\documentclass[aps,prc,amsmath,amssymb,reprint,superscriptaddress]{revtex4-2}

\usepackage{graphicx}
\usepackage{dcolumn}
\usepackage{longtable}
\usepackage{bm}

\begin{document}


\title{Measurement of the production branching ratios following nuclear muon capture for palladium isotopes using the in-beam activation method }


\author{M.~Niikura}
\email{niikura@riken.jp}
\affiliation{Department of Physics, Graduate School of Science, The University of Tokyo, Hongo, Bunkyo, 113-0033 Tokyo, Japan.}
\affiliation{RIKEN Nishina Center, RIKEN, Wako, 351-0198 Saitama, Japan.}
\author{T.~Y.~Saito}
\affiliation{Department of Physics, Graduate School of Science, The University of Tokyo, Hongo, Bunkyo, 113-0033 Tokyo, Japan.}
\affiliation{Atomic, Molecular, and Optical Physics Laboratory, RIKEN, Wako, 351-0198 Saitama, Japan.}
\affiliation{Center for Nuclear Study (CNS), Graduate School of Science, The University of Tokyo, Wako, 351-0198 Saitama, Japan.}
\author{T.~Matsuzaki}
\affiliation{RIKEN Nishina Center, RIKEN, Wako, 351-0198 Saitama, Japan.}
\author{K.~Ishida}
\affiliation{RIKEN Nishina Center, RIKEN, Wako, 351-0198 Saitama, Japan.}
\affiliation{High Energy Accelerator Research Organization (KEK), Tsukuba, 305-0801 Ibaraki, Japan.}
\author{A.~Hillier}
\affiliation{SFTC Rutherford Appleton Laboratory, Didcot, Oxfordshire OX11 0QX, United Kingdom.}


\date{\today}

\begin{abstract}
\begin{description}
\item[Background]
The energy distribution of excited states populated by the nuclear muon capture reaction can facilitate an understanding of the reaction mechanism; however, experimental data are fairly sparse.
\item[Purpose]
We developed a novel method, called the in-beam activation method, to measure the production probability of residual nuclei by muon capture.
For the first application of the new method, we have measured the muon-induced activation of five isotopically-enriched palladium targets.
\item[Methods]
The experiment was conducted at the RIKEN-RAL muon facility of the Rutherford Appleton Laboratory in the UK.
The pulsed muon beam impinged on the palladium targets, and $\gamma$ rays from the $\beta$ and isomeric decays from the reaction residues were measured using high-purity germanium detectors in both the in-beam and offline setups.
\item[Results]
The production branching ratios of the residual nuclei of muon capture for five palladium isotopes with mass numbers $A=$ 104, 105, 106, 108, and 110 were obtained.
The results were compared with a model calculation using the particle and heavy ion transport system (PHITS) code.
The model calculation well reproduces the experimental data.
\item[Conclusion]
For the first time, this study provides experimental data on the distribution of production branching ratios without any theoretical estimation or assumptions in the interpretation of the data analysis.
\end{description}
\end{abstract}


\maketitle


\section{Introduction}
\label{sec:introduction}

\begin{figure}
    \centering
    \includegraphics[width=\linewidth]{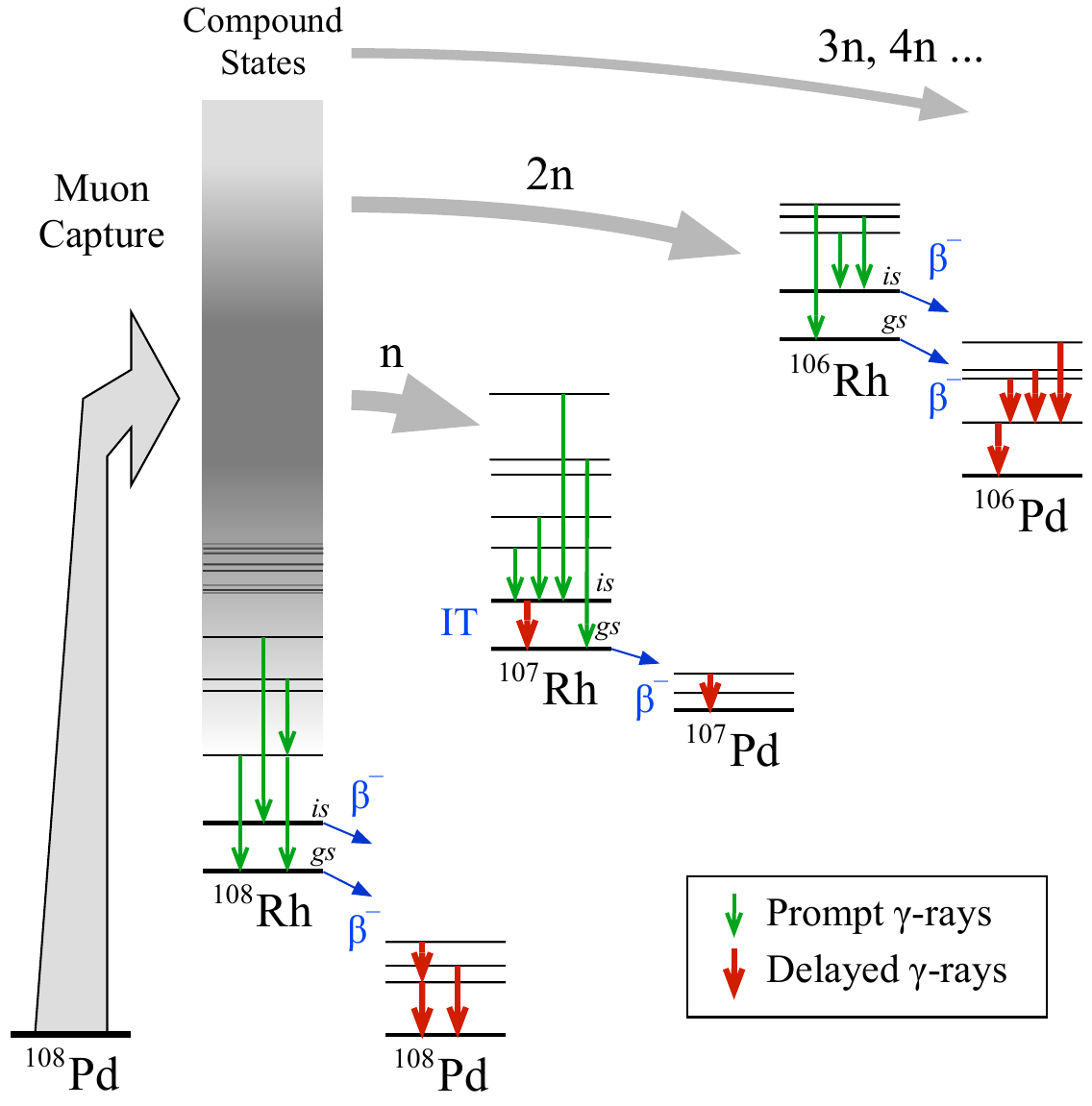}
    \caption{(Color online) Schematic representation of the muon capture process for $^{108}$Pd.
    Muon capture of $^{108}$Pd produces excited (compound) states of $^{108}$Rh$^*$ at around 10--50 MeV.
    After prompt particle emissions and $\gamma$-ray transitions, residual nuclei decay via isomeric transition (IT) and $\beta^{-}$ decays, which are delayed events with typical half-lives of more than a few seconds.
    In the present activation measurement, only delayed $\gamma$-rays are measured.
    See text for details.}
    \label{fig:level}
\end{figure}

The nuclear muon capture reaction~\cite{Measday2001-hi} is the capture of a negative muon by a proton via a weak interaction from the 1s state of the muonic atom, which is expressed as follows:
\begin{equation}
    \mu^- + p \rightarrow n + \nu_\mu.
    \label{eq:capture_mup}
\end{equation}
This reaction is analogous to the electron capture reaction; the crucial difference between electron and muon capture is in their energy and momentum transfer.
The energy released by muon capture is 104.3 MeV, which corresponds primarily to the mass of the muon.
If the proton is at rest, as expressed in Eq.~(\ref{eq:capture_mup}), the recoiling neutron takes only 5.2 MeV of kinetic energy, whereas the neutrino takes away 99.1 MeV.
When muon capture occurs in the nucleus of $(A,Z)$, where $A$ is the mass of the nucleus and $Z$ is the element number, the reaction produces a compound nucleus of $(A,Z-1)^*$ as follows:
\begin{equation}
    \mu^- + (A,Z) \rightarrow (A,Z-1)^* + \nu_\mu.
    \label{eq:capture_munucl}
\end{equation}
Figure~\ref{fig:level} shows a schematic representation of the muon capture process for $^{108}$Pd.
Because the nucleus is a many-body system, the excitation energy of the compound nucleus is expected to be distributed around 10--50 MeV.
The energy distribution of excited states populated by muon capture can facilitate an understanding of the reaction mechanism; however, experimental data are sparse and require improvement.
Because the emitted neutrinos are barely detected, missing mass spectroscopy can not be applied to obtain the excitation energy distribution.
Thus, the excited states of the compound nucleus have been investigated by measuring other emitted particles, such as neutrons, $\gamma$ rays, protons, and alphas, and the production branching ratios of the residual nuclei.

In medium-heavy nuclei, the particles emitted from muon capture are primarily neutrons because the emission of charged particles is suppressed by the Coulomb barrier.
The energy spectra of neutrons have been measured for the heavy nuclei of Tl, Pb, and Bi~\cite{Schroder1974-jb}.
The low-energy component of the neutron energy spectrum below 5 MeV can be explained by the statistical evaporation from the compound nucleus; however, the spectrum extends to higher energies.
The high-energy component of the neutron energy spectrum is interpreted as due to direct and preequilibrium processes, in which the neutron is emitted immediately at the time of muon capture before reaching the thermal equilibrium of the compound states.
Neutron multiplicity has also been measured in the past using a large liquid scintillator tank~\cite{Macdonald1965-tw}.
The scintillator tank has a high neutron detection efficiency of 54.5\%; however, even this is not sufficiently high to obtain a reliable multiplicity distribution because of the large error propagation from the probability of the high multiplicity events in the unfolding procedure~\cite{Measday2001-hi}.

The production branching ratio for muon capture can be deduced from the prompt $\gamma$-ray measurements~\cite{Backenstoss1971-qj, Johnson1996-he, Gorringe1999-je, Stocki2002-vn, Measday2006-ti, Measday2007-ed, Zinatulina2019-nv}.
Muon capture populates the excited state of the reaction residues and decays with the emission of $\gamma$ rays.
By measuring the characteristic $\gamma$-ray transitions to the ground state, one can determine the number of residual nuclei produced.
Although most of the transitions (more than 90\% of the total yields) were observed in some cases, there could be missing yields because of the existence of weak transitions, unidentified $\gamma$-ray energies, and direct population of the ground state after particle emissions.
Hence, the production branching ratio deduced from prompt $\gamma$-ray measurements always yields a lower limit.

The activation method is the most reliable and sensitive technique for determining the production rate of radioactive nuclei by beam irradiation.
The production branching ratios of muon capture have been measured for several nuclei~\cite{Bunatyan1969-sp, Bunatyan1970-vi, bunatyan_1972, Heusser1972-oa, Wyttenbach1978-ss, Heisinger2002-dh} by the activation method.
In ordinary activation measurements, only production ratios of long-lived radioactive isotopes can be obtained because the decay measurements normally take place separately at the time and location of beam irradiation to avoid the beam background.
Therefore, in this study, to measure short-lived states by the activation method, we developed a novel method called the in-beam activation method.
In low-energy muon beam facilities based on the synchrotron, such as RAL-ISIS and J-PARC MLF, the muon beam has a pulsed time structure, in which the muon beam has a pulse width of a few hundred nanoseconds and the interpulse period is a few tens of milliseconds.
Because there is no beam background during the interpulse period, it is ideal for decay measurements.
In the in-beam activation method, decaying $\gamma$ rays are measured simultaneously with beam irradiation by exploiting the time structure of the pulsed muon beam.
The combination of in-beam and ordinary offline activation methods enables the measurement of most of the $\beta$-decaying states with a wide range of half-lives from a few milliseconds to years.

For the first application of the in-beam activation method, we have measured the activation of five isotopically enriched palladium targets: $^{104,105,106,108,110}$Pd.
The choice of the palladium targets is based on the available enriched targets with even proton numbers (even-$Z$) in medium-heavy nuclei.
Neutron evaporation is the primary decay process of compound nuclei produced by muon capture for the medium-heavy nuclei, and the majority of the reaction residues are $Z-1$ isotopes.
Because stable isotopes cannot be measured using the activation method and odd-$Z$ nuclei have fewer stable isotopes, the even-$Z$ target is ideal for the measurement of reaction residues using the activation method.

Here, we define the notation used in the present study:
the muon capture reaction on, for example, $^{108}$Pd produces excited states of $^{108}$Rh: $^{108}\mathrm{Pd}(\mu,\nu_\mu){}^{108}\mathrm{Rh}^*$.
We refer to this $^{108}$Rh$^*$ as the compound nucleus, although part of muon capture undergoes direct and preequilibrium processes.
The reaction channels are named based on the number of protons and neutrons emitted from the compound nucleus; for example, the production of $^{108}$Rh and $^{106}$Ru from muon capture of $^{108}$Pd are called 0p0n and 1p1n channels, respectively.
Because reaction products from charged particle emissions are rarely observed, the number of proton emissions (0p) is sometimes omitted from this notation.
There are several isomeric states of the rhodium isotopes.
For instance, $^{108}$Rh has two $\beta$-decaying states, namely ground and isomeric states, and they are labeled as \textit{''gs``} or $^{108g}$Rh and \textit{''is``} or $^{108m}$Rh, respectively.
$\Delta X$ denotes the uncertainty of parameter $X$.

This paper is organized as follows: in Sect.~\ref{sec:experiment}, the experimental setup at the RIKEN-RAL muon facility is described.
The general analysis procedure for the in-beam activation method is explained in Sect.~\ref{sec:analysis} and the results for each target together with a detailed data treatment are presented in Sect.~\ref{sec:result}.
In Sect.~\ref{sec:discussion}, the obtained production branching ratios and features of the newly proposed in-beam activation method are discussed.
Finally, we conclude the present study in Sect.~\ref{sec:conclusion}.

\section{Experiment}
\label{sec:experiment}

\begin{table*}
    \caption{Isotope composition of the enriched Pd targets~\cite{Saito2022-ju}.}
    \label{tab:target_composition}
    \begin{ruledtabular}
    \begin{tabular}{c D{.}{.}{-1} D{.}{.}{-1} D{.}{.}{-1} D{.}{.}{-1} D{.}{.}{-1} D{.}{.}{-1} D{.}{.}{-1}}
    Target & \multicolumn{1}{l}{Chemical} &  \multicolumn{6}{c}{Abundance} \\
    & \multicolumn{1}{l}{purity (\%)} &
    \multicolumn{1}{c}{$^{102}$Pd} &
    \multicolumn{1}{c}{$^{104}$Pd} &
    \multicolumn{1}{c}{$^{105}$Pd} &
    \multicolumn{1}{c}{$^{106}$Pd} &
    \multicolumn{1}{c}{$^{108}$Pd} &
    \multicolumn{1}{c}{$^{110}$Pd} \\ \hline
    $^{104}\mathrm{Pd}$ & 99.97 & <0.02    & 98.4(1)  & 1.05(5) & 0.35(3)  & 0.18(2)   & <0.05    \\
    $^{105}\mathrm{Pd}$ & 99.97 & 0.033(6) & 0.236(4) & 97.9(7) & 1.407(8) & 0.311(4)  & 0.112(2) \\
    $^{106}\mathrm{Pd}$ & 99.97 & <0.03    & 0.06     & 0.68    & 98.4(2)  & 0.8       & 0.06     \\
    $^{108}\mathrm{Pd}$ & 99.97 & <0.02    & 4.8(1)   & 0.15(3) & 0.90(5)  & 93.80(15) & 0.30(3)  \\
    $^{110}\mathrm{Pd}$ & 99.99 & <0.05    & 0.1      & 0.35    & 0.5      & 0.7       & 98.3(2)  \\
    \end{tabular}
    \end{ruledtabular}
\end{table*}

\begin{table*}
    \caption{Summary of targets, beam momenta ($p_\mathrm{beam}$), in-beam measurement (beam irradiation) time and off-beam (after beam irradiation) measurement time, number of muons irradiated ($N_\mu$), stopping efficiency ($\epsilon_\mathrm{stop}$), and the number of muons captured ($N_\mathrm{cap}$). See the text for further details.}
    \label{tab:targets}
    \begin{ruledtabular}
    \begin{tabular}{cc D{.}{.}{-1} D{.}{.}{-1} D{t}{}{-1} D{t}{\times}{-1} D{.}{.}{3} D{t}{\times}{-1} }
        \multicolumn{2}{c}{Target} &
        \multicolumn{1}{c}{$p_\mathrm{beam}$ (MeV/c)} &
        \multicolumn{2}{c}{Measurement time (h)} &
        \multicolumn{1}{c}{$N_\mu$} &
        \multicolumn{1}{c}{$\epsilon_\mathrm{stop}$} &
        \multicolumn{1}{c}{$N_\mathrm{cap}$\footnotemark[1]} \\
        &&& \multicolumn{1}{c}{in-beam} & \multicolumn{1}{c}{off-beam} \\
        \hline
        $^{104}$Pd & metal disk     & 33.9(1) & 7.1  & 167t\footnotemark[2] & 1.11t10^7 & 0.911(27) & 1.01t10^7\\
        $^{105}$Pd & powder in case & 34.9(1) & 17.1 & 40t\footnotemark[3]  & 3.10t10^7 & 0.764(31) & 2.37t10^7\\
        $^{106}$Pd & powder in case & 34.9(1) & 17.8 & 36t\footnotemark[4]  & 3.10t10^7 & 0.957(10) & 2.97t10^7\\
        $^{108}$Pd & metal disk     & 33.9(1) & 8.8  & 30t\footnotemark[3]  & 1.35t10^7 & 0.812(10) & 1.12t10^7\\
        $^{110}$Pd & metal disk     & 33.9(1) & 10.1 & -t    & 1.55t10^7 & 0.810(24) & 1.26t10^7\\
    \end{tabular}
    \footnotetext[1]{$N_\mathrm{cap}$ has a systematic uncertainty from muon beam intensity calibration (2\%), $P_\mathrm{cap}$ (1\%) and $\epsilon_\mathrm{stop}$ (1--4\%).}
    \footnotetext[2]{Decay for $^{104}$Pd was measured at the UT offline setup.}
    \footnotetext[3]{Decays for $^{105}$Pd and $^{108}$Pd were measured at the RAL offline setup.}
    \footnotetext[4]{Decay for $^{106}$Pd was measured at the in-beam setup (12 hours) and the RAL offline setup (24 hours).}
    \end{ruledtabular}
\end{table*}

The experiment was conducted at the RIKEN-RAL muon facility of the Rutherford Appleton Laboratory (RAL) in the United Kingdom~\cite{Matsuzaki2001-ty}.
A proton beam of 800~MeV from the ISIS synchrotron irradiated an intermediate graphite target at Target Station 1 (TS1) to produce pions.
The primary beam had a double-pulse structure with a 50Hz repetition rate, and four out of five pulses were sent to TS1.
The negative muon beam, which is a decay product of the negative pions, was transported and momentum-analyzed through the RIKEN-RAL beamline and delivered to Port-1, where the experimental apparatus was installed.

\begin{figure}
    \centering
    \includegraphics[width=8.6cm]{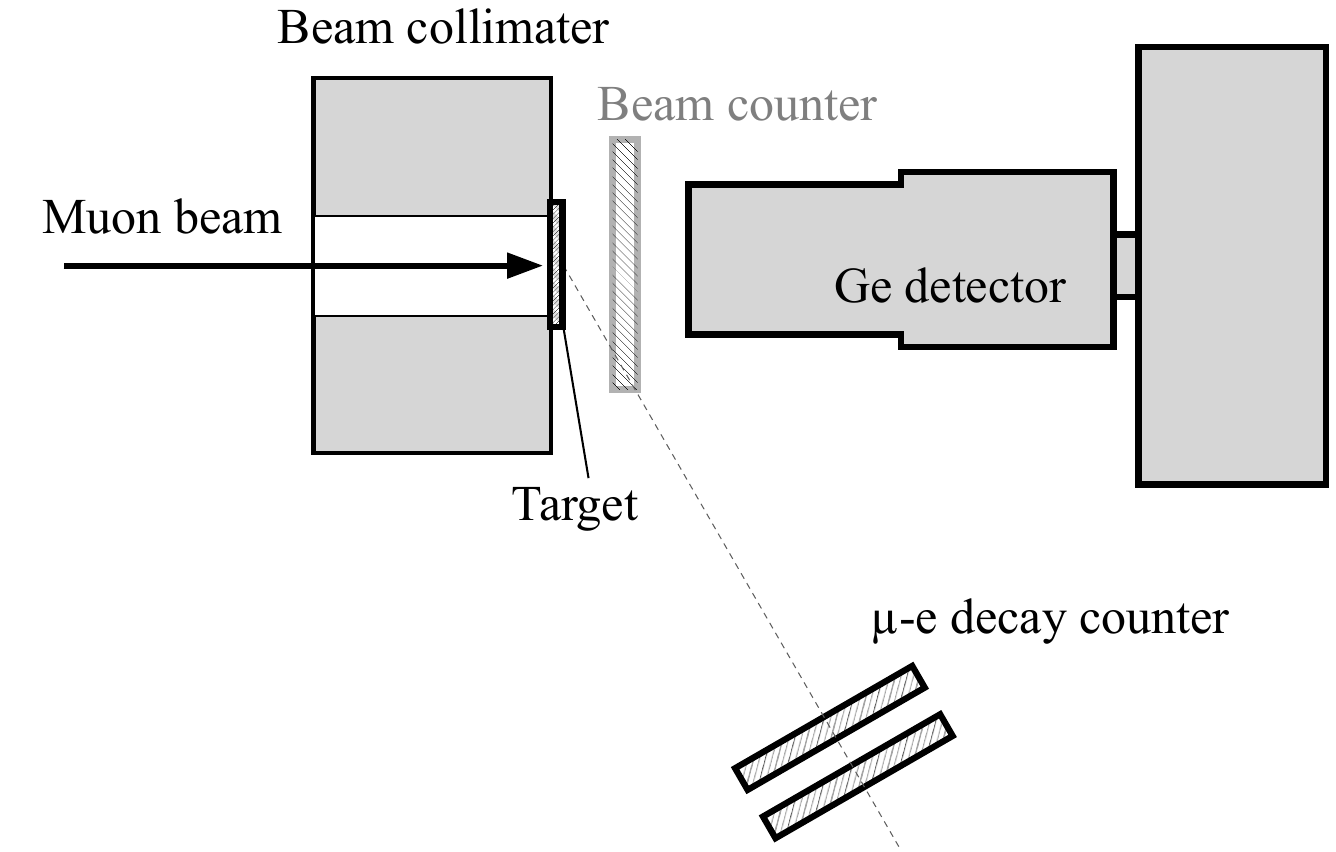}
    \caption{Schematic of the in-beam activation setup at Port-1 at the RIKEN-RAL muon facility (not to scale). The negative muon beam is derived from the left side of the figure. The beam passes through the beam collimator and stops at the target attached to the downstream side surface of the collimator. The beam counter is installed only prior to the activation runs to measure the beam intensity and is removed during the activation measurements. A germanium detector is placed downstream of the target to detect the $\beta$-delayed $\gamma$ rays. The $\mu$-$e$ decay counter consisting of two plastic scintillators is used to monitor the beam status by detecting electrons from the muon decay.}
    \label{fig:setup}
\end{figure}

Figure~\ref{fig:setup} shows a schematic of the in-beam activation setup (referred to as in-beam setup, hereafter).
The muon beam was passed through a beam collimator of a size 100$\times$100$\times$60~mm$^3$ made of acrylic with a hole diameter of 14~mm.
The palladium targets were attached to the downstream side of the collimator and irradiated with the muon beam.
The activation of five isotopically enriched metallic palladium targets ($^{104,105,106,108,110}$Pd) was measured in the present experiment.
The targets were also used in our previous muonic X-ray measurements, and a list of the targets and their isotopic compositions is shown in Table~\ref{tab:target_composition}~\cite{Saito2022-ju}.
The $^{104}$Pd, $^{108}$Pd, and $^{110}$Pd targets were metal discs with a diameter of 15 mm and a thickness of 0.5 mm.
The $^{105}$Pd and $^{106}$Pd targets were metal powders encapsulated in a graphite case with a thickness of 1 mm on each side.
The effective sizes of the powder targets were 20 mm in diameter and 2.2 mm in thickness for the $^{105}$Pd target and 15 mm in diameter and 2.3 mm in thickness for the $^{106}$Pd target, respectively.
A beam counter was used only prior to the activation measurement without the target to measure the number of muons in the beam pulse, which were removed during the activation measurements.
The beam counter consisted of a plastic scintillator of a size 50$\times$50$\times$5~mm$^3$.
The $\beta$-delayed $\gamma$ rays from the activated targets were detected using a high-purity n-type coaxial germanium detector with 26.6\% relative efficiency (ORTEC GMX-20P4).
The distance between the target and the front surface of the germanium detector was 45~mm.
At this close distance, muonic X rays and prompt $\gamma$ rays from muon capture cannot be measured because of a pile-up of the output signal; only delayed $\gamma$ rays can be measured during the interpulse period of the pulsed muon beam.
A $\mu$-$e$ decay counter consisting of two plastic scintillators was placed at 145 mm and 245 mm from the target and 60 degrees relative to the beam direction.
It was used to monitor the beam status (beam on/off) during the experiment by detecting the decay electrons of the muons stopped at the target and beam collimator.

The signals from the detectors were processed using two waveform digitizers with a 500-MHz sampling rate and 14-bit resolution (CAEN V1730B).
The energy and time-stamp of the $\gamma$ rays were taken by the digitizer with the Digital Pulse Processing for Pulse Height Analysis (DPP-PHA) firmware under a self-trigger condition.
The dynamic range of the measured $\gamma$-ray energy is set to 40--1800~keV.
The typical count rate of the germanium detector was approximately 100 counts per second (cps) during beam irradiation and approximately 50 cps without the beam (environmental background).
The time-stamp of the pulsed beam and signal waveform of the plastic scintillators were recorded using the digitizer with the WaveDump firmware with a 50-Hz trigger condition from the accelerator.
The count rate of the $\mu$-$e$ decay counter, defined as the coincidence of the two plastic scintillator signals, was recorded using a scaler every second.

The muon beam momentum ($p_\mathrm{beam}$) was chosen to stop most of the muon beam at the target and prevent it from hitting the germanium detector. 
By considering the graphite case thickness for the powder targets, we set $p_\mathrm{beam}$ to 33.9(1)~MeV/c for the $^{104,108,110}$Pd disk targets and 34.9(1)~MeV/c for the $^{105,106}$Pd powder targets.

The $\beta$-decays of the reaction products with longer half-lives, for example, $^{101m}$Rh decay ($T_{1/2}=4.34$~days), $^{102g}$Rh decay ($T_{1/2}=207.3$~days), and $^{105g}$Rh decay ($T_{1/2}=35.3$~hours), were measured using two offline setups, located outside the experimental area.
The first offline setup was located next to Port-1 at RAL (called RAL offline setup, hereafter).
The setup consisted of a high-purity p-type coaxial germanium detector with 8\% relative efficiency (ORTEC GEM-S5020P4-B) with lead shields.
The typical count rate of the detector was approximately 10 cps.
The second offline setup was located at the University of Tokyo in Japan (called UT offline setup, hereafter) and consisted of a high-purity p-type coaxial germanium detector with 30.2\% relative efficiency (ORTEC GEM-25195) and multi-layer shields made of lead and copper for ultra-low background measurements.
The UT offline setup was used only for the decay measurement of the $^{104}$Pd target.
The activated target was placed in front of the detector at distances of 25~mm (RAL offline setup) and 2~mm (UT offline setup).
A conventional shaping amplifier (ORTEC 572A) and a multi-channel analyzer (MCA) were used to obtain the energy spectra of the germanium detectors in the offline setups.

The beam momenta ($p_\mathrm{beam}$), in-beam measurement (beam irradiation) time and off-beam measurement (after beam irradiation) time, and number of muons irradiated ($N_\mu$) for each target run are summarized in Table~\ref{tab:targets}.

\section{Data analysis}
\label{sec:analysis}

The production branching ratios of reaction products for five isotopically enriched targets ($b'$) were deduced from the number of observed $\gamma$ rays from $\beta$ and isomeric decays ($N_\gamma$) using the following formula:
\begin{align}
    b' = \frac{N_\gamma/(\epsilon_\gamma \epsilon_\mathrm{LT})}{N_\mathrm{cap} P_\mathrm{decay} I_\gamma},
    \label{eq:branch}
\end{align}
where $\epsilon_\gamma$ is the $\gamma$-ray detection efficiency of the germanium detectors, $\epsilon_\mathrm{LT}$ is the analysis live-time ratio, $N_\mathrm{cap}$ is the number of capture reactions, $P_\mathrm{decay}$ is the decay probability during the measurement period, and $I_\gamma$ is the $\gamma$-ray intensity per decay of the parent nuclei ($\beta$ or isomeric decay).
The extraction of these parameters is described in this section.

\begin{figure}
    \centering
    \includegraphics[width=8.6cm]{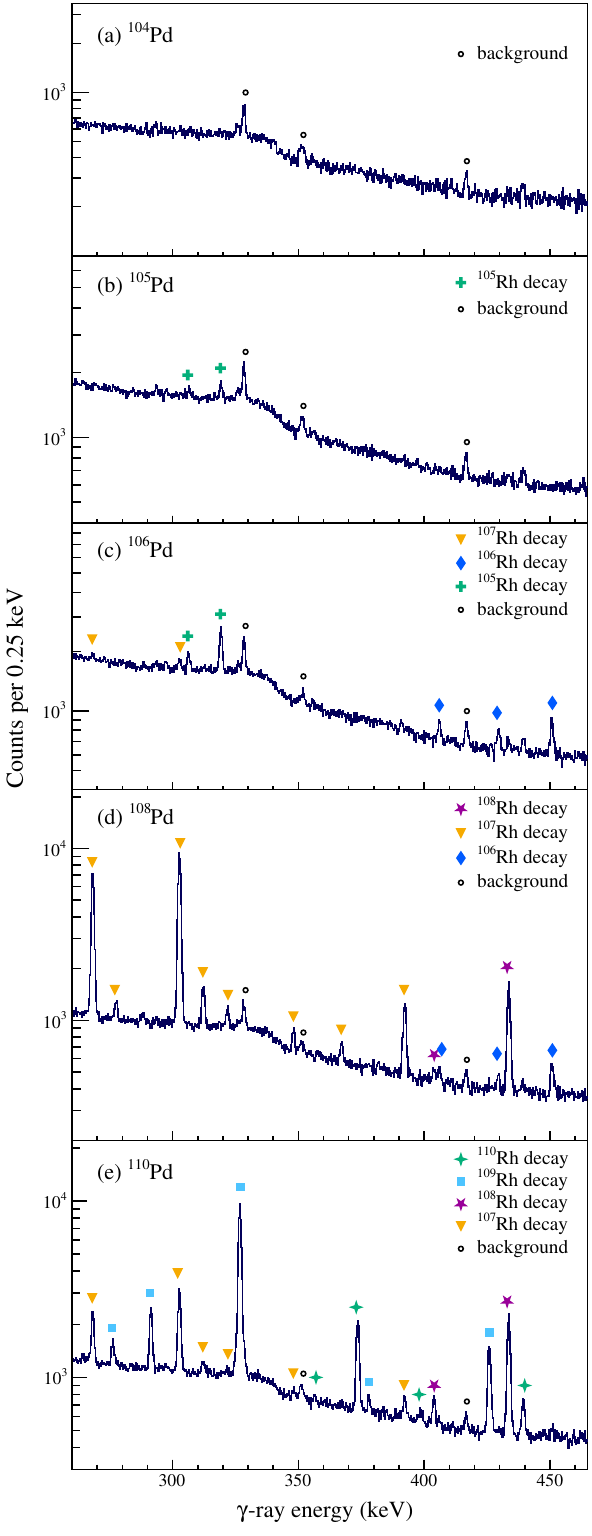}
    \caption{(Color online) Part of the delayed $\gamma$-ray spectra of the $^{104,105,106,108,110}$Pd activation.
    Peaks marked with closed symbols are from $\beta$- or isomeric-decay $\gamma$ rays, and those with open circles are backgrounds.}
    \label{fig:gamma}
\end{figure}

Figure~\ref{fig:gamma} shows a part of the in-beam $\gamma$-ray spectra of $^{104,105,106,108,110}$Pd activation.
$\gamma$-ray peaks from $\beta$ and isomeric decays of the radioactive products of muon capture were observed in the spectra.
The number of each $\gamma$-peak ($N_\gamma$) was obtained by fitting the peaks with a Gaussian function and a linear background term.

The $\gamma$-ray detection efficiencies of the germanium detectors ($\epsilon_\gamma$) in the in-beam and RAL offline setups were measured using the standard $\gamma$-ray sources of $^{22}$Na, $^{60}$Co, $^{133}$Ba, and $^{137}$Cs.
$\epsilon_\gamma$ for the UT offline setup was estimated by the Monte Carlo simulation using the \texttt{Geant4} toolkit~\cite{geant4-1,geant4-2,geant4-3}.
The absolute efficiencies are, for example, 1.6\% for the in-beam setup, 1.9\% for the RAL offline setup, and 14\% for the UT offline setup for the 302.9-keV $\gamma$ ray of the $^{133}$Ba decay, respectively.
The absolute detection efficiency had a 3\% systematic uncertainty originating from the uncertainty of source activities provided by the manufacturer.

The acquisition live-time ratio was almost unity ($>99.99\%$) because of the dead-time less feature of the waveform digitizers used in the in-beam setup.
For in-beam measurements, the analysis live-time ratio should be considered. 
Because the germanium detector was placed so close to the target at zero degrees, muonic X-rays and subsequent $\gamma$ rays from muon capture as well as the electron contaminant in the beam hit the detector at the prompt timing of the beam arrival.
These photons and electrons caused pile-up and saturation of the pre-amplifier output signal of the germanium detector.
To eliminate the inefficient time by the pile-up, 0.5~ms after beam arrival was considered as a dead time ($T_d$) and excluded in the analysis hereafter.
The analysis live-time ratio ($\epsilon_\mathrm{LT}$) is expressed as follows:
\begin{equation}
    \epsilon_\mathrm{LT} = \frac{\int_{T_d}^{T_p} \exp(-\lambda t)dt}{\int_0^{T_p}\exp(-\lambda t)dt},
    \label{eq:lifetime}
\end{equation}
where $T_p$ is the beam period of 20~ms for the ISIS synchrotron (50Hz frequency) and $\lambda$ is the decay constant of each reaction product ($\lambda = \ln(2)/T_{1/2}$, where $T_{1/2}$ is a half-life of the reaction product).
All observed states in the present experiment had considerably longer half-lives than the beam period ($T_{1/2}\gg20$~ms), and  $\epsilon_\mathrm{LT}=19.5/20.0$ as $\lambda\rightarrow0$ limit of Eq.~(\ref{eq:lifetime}) was used in the analysis.
Correspondingly, decaying $\gamma$ rays from a state with a half-life shorter than $T_d$ could not be observed, because of the small value of $\epsilon_\mathrm{LT}$.
For the offline setups, the acquisition live-time ratio was almost unity ($>99.99\%$) owing to the low count rate.

The absolute intensity of each measured $\gamma$-ray per $\beta$ decay ($I_\gamma$) was obtained from Evaluated and Compiled Nuclear Structure Data (ENSDF)~\cite{nndc_100,nndc_101,nndc_102,nndc_104,nndc_105,nndc_106,nndc_107,nndc_108,nndc_109,nndc_110}.
There was some incompleteness in the database, and the evaluation of these data is explained in the next section.

The number of muon captures ($N_\mathrm{cap}$) was estimated using two independent methods.
The first method is based on the fact that the sum of the production branching ratio is 100\% ($\sum_\mathrm{nucl}b'=1$), \textit{i.e.} the sum of the number of produced nuclei ($N_\mathrm{nucl}$) should be the total number of the capture reaction, which is expressed as follows:
\begin{equation}
    N_\mathrm{cap1} = \sum_\mathrm{nucl} N_\mathrm{nucl} + M,
\end{equation}
where
\begin{equation}
    N_\mathrm{nucl} = \frac{N_\gamma/(\epsilon_\gamma \epsilon_\mathrm{LT})}{P_\mathrm{decay} I_\gamma}
\end{equation}
and $M$ is the sum of the missing products in the present setup.
The missing products originate from the production of stable isotopes (e.g., $^{103}$Rh and $^{102,104}$Ru), weak $\gamma$-ray emissions via $\beta$ decay because of their long half-lives ($^{101g,102m}$Rh), or small $I_\gamma$ values ($^{104g}$Rh, $^{106}$Ru).
Because $M$ was barely estimated in most cases, the first method was used only to constrain the lower limit of $N_\mathrm{cap}$ under the condition $M=0$.

The second method involves direct counting of the muon beam.
The total number of muon captures ($N_\mathrm{cap2}$) was deduced from the total number of irradiating muons ($N_\mu$) as follows:
\begin{equation}
    N_\mathrm{cap2} = N_\mu \epsilon_\mathrm{stop} P_\mathrm{cap},
\end{equation}
where $\epsilon_\mathrm{stop}$ is the stopping efficiency of the beam in the targets and $P_\mathrm{cap}$ is the muon capture probability of the 1s state of the muonic atom.
$\epsilon_\mathrm{stop}$ was estimated with the Monte Carlo simulation using the \texttt{G4BeamLine} code~\cite{g4beamline}.
In the simulation, the measured geometry of the experimental setup, beam momenta ($p_\mathrm{beam}$) and their distribution (3.1(3)\%), measured target shape and thickness, and energy loss at the exit window of the beamline were considered.
The uncertainty of $\epsilon_\mathrm{stop}$ is estimated from that of beam momenta and their distribution, and thicknesses of the target and the exit window.
$\epsilon_\mathrm{stop}$ for each target run are listed in Table~\ref{tab:targets}.
Muon capture probability ($P_\mathrm{cap}$) was deduced from the total capture rate of natural palladium ($\Lambda_C=10.00(7)$ $\mu$s$^{-1}$)~\cite{Suzuki1987-aq}.
$P_\mathrm{cap}$ was calculated using the following formula:
\begin{equation}
    P_\mathrm{cap} = \frac{\Lambda_C}{Q/\tau_{\mu^+}+\Lambda_C},
\end{equation}
where $\tau_\mu^+$ is the lifetime of the positive muon of 2.196811(22)~$\mu$sec~\cite{Particle_Data_Group2022-lz} and $Q$ is the Huff factor of 0.927~\cite{Suzuki1987-aq,Huff1961-ur}.
The deduced value of $P_\mathrm{cap}=0.960(10)$ was used for all enriched targets, in which the quoted uncertainty included $\Delta\Lambda_C$ and the isotope dependence of $\Lambda_C$ as a systematic uncertainty by taking the standard deviation of the measured values for each enriched target in Ref.~\cite{Saito2022-ju}.

\begin{figure}
    \centering
    \includegraphics[width=8.6cm]{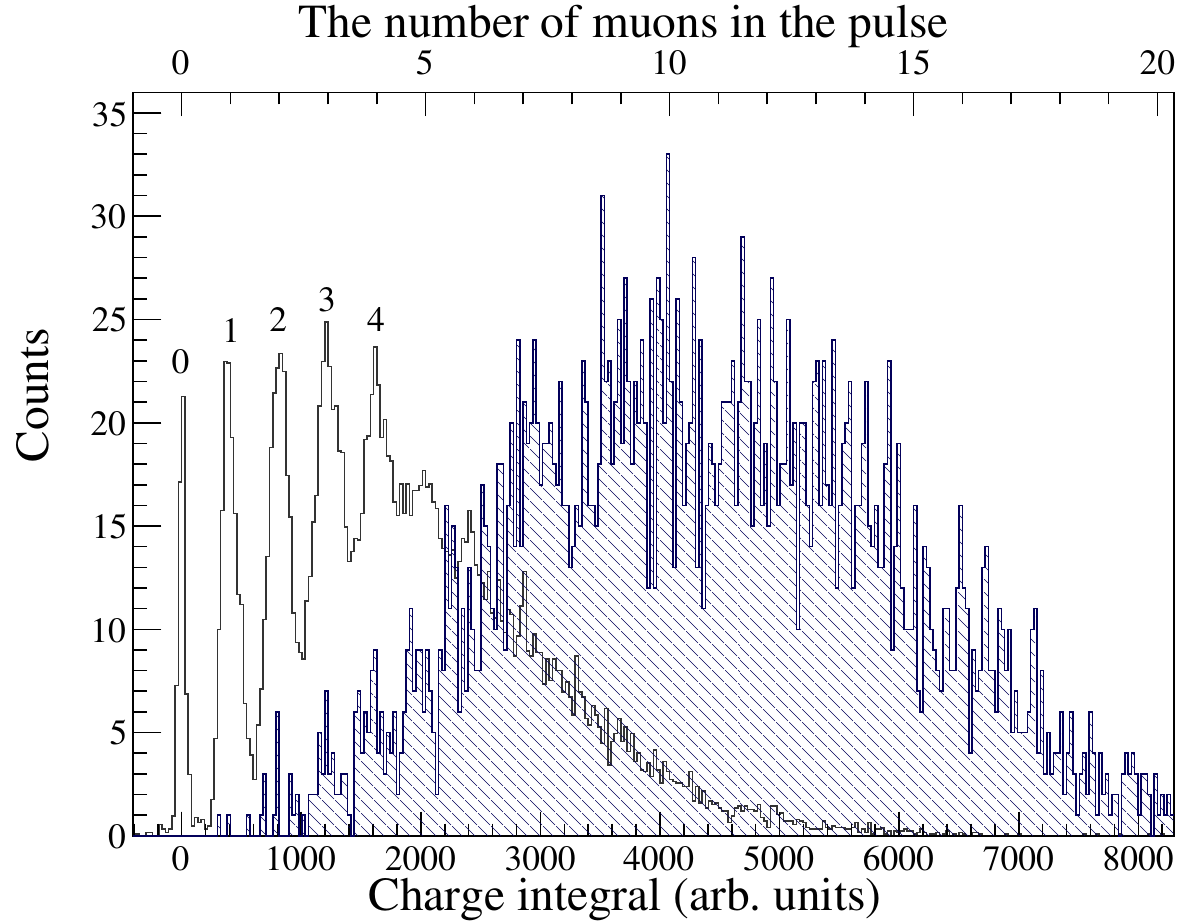}
    \caption{(Color online) Number of muons in the double pulse ($n_\mu$) measured by the beam counter for 33.9-MeV/c muon beam.
    The bottom horizontal axis represents the charge integral of the beam counter signal, and the top horizontal axis is calibrated to $n_\mu$ using a low-intensity muon beam (open spectrum).
    The shaded histogram shows a $n_\mu$ distribution with the same beamline setting as the activation measurement.}
    \label{fig:beam}
\end{figure}
The muon beam intensity was measured using the beam counter prior to the activation measurements.
Figure~\ref{fig:beam} shows the spectra of the charge integral of the beam counter signal for the 33.9-MeV/c muon beam.
The charge integration of the scintillator signal is proportional to the number of muons in the double pulses ($n_\mu$).
The open spectrum in the figure shows the charge integration for a low-intensity muon beam.
The spectrum shows discrete peaks corresponding to the number of muons ($n_\mu = 0,1,2,3...$) and was used for calibration from charge integration to $n_\mu$.
The calibrated value of $n_\mu$ is shown at the top of the horizontal axis in the figure.
The shaded spectrum shows the beam intensity with the beamline setting for the 33.9-MeV/c muon beam used for the activation measurement.
The average number of muons in each double pulse was 11.2(3) for a 169.2-$\mu$A primary beam current.
Because the beam counter was removed during the activation measurements, the muon beam intensity was monitored using the proton beam current from the ISIS synchrotron assuming that the muon beam intensity was proportional to the proton beam current.
Using the actual beam frequency of 40 pulses per second at TS1, the conversion parameters from the proton beam current to the muon beam intensity were 2.65(5) and 3.05(5) particles/s$\cdot\mu$A$^{-1}$ for the 33.9- and 34.9-MeV/c settings, respectively.
The total number of irradiating muons ($N_\mu=\int 40\cdot n_\mu dt$) was then derived from the time integral of the primary beam current with the calibration parameters listed in Table~\ref{tab:targets} for each target run.

\begin{figure}
    \centering
    \includegraphics[width=8.6cm]{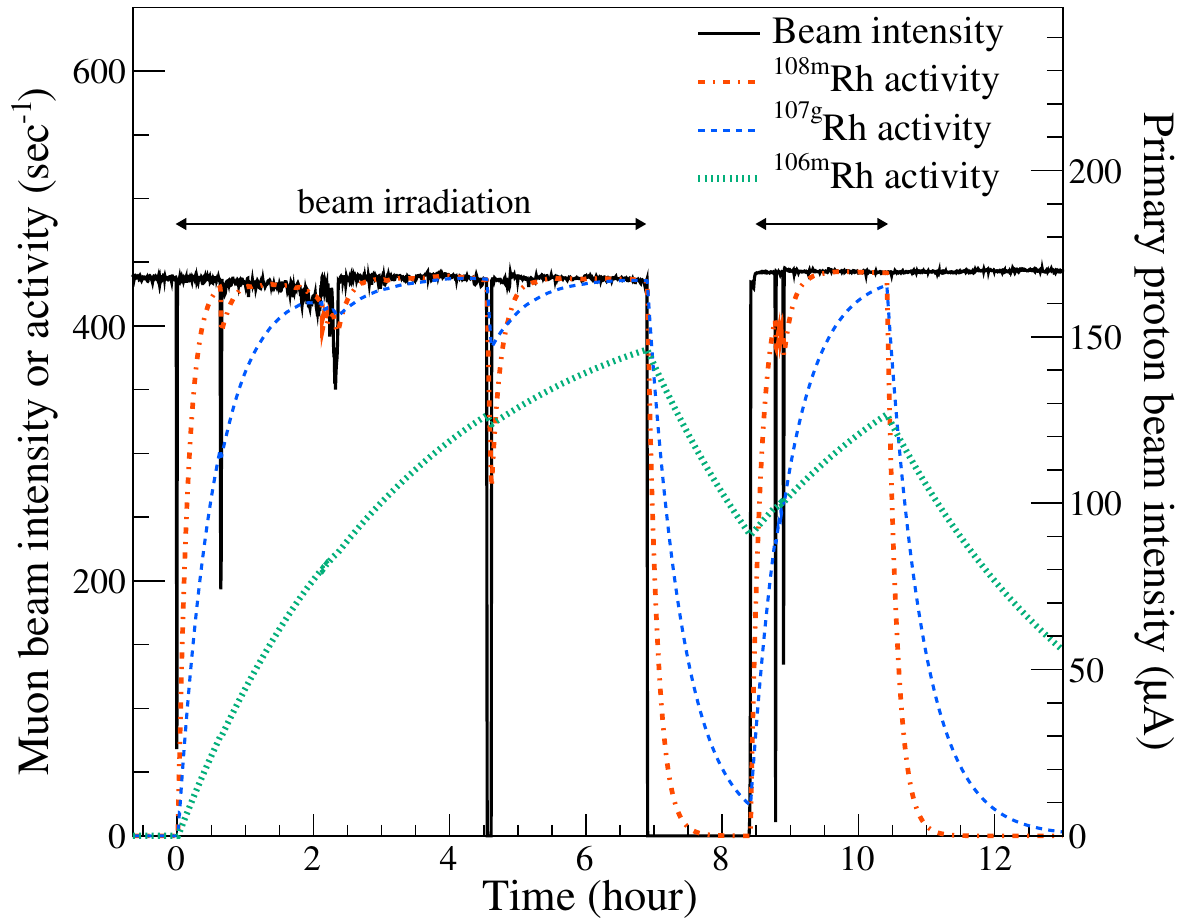}
    \caption{(Color online) Beam intensity and activity curve of the reaction products for $^{108}$Pd activation run.
    The black solid line represents the muon beam intensity (left vertical axis) deduced from the proton beam intensity (right vertical axis) assuming that they are proportional.
    The orange dashed-dotted, blue dashed, and green dotted lines represent the calculated activity curve with half-lives of $^{108m}$Rh ($T_{1/2}=6.0~\mathrm{min}$)~\cite{nndc_108}, $^{107g}$Rh ($T_{1/2}=21.7~\mathrm{min}$)~\cite{nndc_107}, and $^{106m}$Rh ($T_{1/2}=131~\mathrm{min}$)~\cite{nndc_106}, respectively, assuming $b'=1$.}
    \label{fig:decay}
\end{figure}

The differential equation for radioactive decay is expressed as follows:
\begin{equation}
    \frac{dn_\mathrm{nucl}(t)}{dt} = -\lambda n_\mathrm{nucl}(t) + y_\mathrm{nucl}(t),
    \label{eq:differential}
\end{equation}
where $n_\mathrm{nucl}(t)$ is the number of radioactive reaction products and $y_\mathrm{nucl}(t)$ is the production yield of the radioactive state by muon capture.
The decay probability during the measurement time ($P_\mathrm{decay}$) is defined as follows:
\begin{equation}
    P_\mathrm{decay} \equiv \frac{\int \lambda n_\mathrm{nucl}(t)dt}{N_\mu}
    \label{eq:decay}
\end{equation}
under the condition $y_\mathrm{nucl}=40\cdot n_\mu$, \textit{i.e.} assuming that the irradiating muon produces a given nucleus with $b'=1$.
The integral range of the numerator in Eq.~(\ref{eq:decay}) represents the measurement time.
The measurement time can differ from the irradiation time, and Eq.~(\ref{eq:decay}) is applicable to both the in-beam and offline activation measurements.
For example, if the beam intensity is constant and the measurement time is the same as the irradiation time, we can analytically calculate $P_\mathrm{decay}$ during in-beam activation as follows:
\begin{equation}
    P_\mathrm{decay} = \frac{\int_{t_\mathrm{start}}^{t_\mathrm{stop}} (1-\exp(-\lambda t))dt}{t_\mathrm{stop}-t_\mathrm{start}},
    \label{eq:decayinbeam}
\end{equation}
where $t_\mathrm{start}$ and $t_\mathrm{stop}$ are the start and stop timings of the measurements, respectively.
The time origin of Eq.~(\ref{eq:decayinbeam}) is the start timing of irradiation and $t_\mathrm{start} \geqq 0$.
For the offline measurements,
\begin{equation}
    P_\mathrm{decay} = \frac{\int_{t_\mathrm{start}}^{t_\mathrm{stop}} \lambda \exp(-\lambda n_\mathrm{nucl}(t)) dt}{N_\mu}.
    \label{eq:decayoffline}
\end{equation}
The uncertainty of $T_{1/2}$ reflects that of $P_\mathrm{decay}$, which was negligible (less than 0.1\%).
In the actual experiment, the beam had a fluctuating intensity and was sometimes interrupted during the measurement.
Therefore, $P_\mathrm{decay}$ was deduced through numerical calculations.
Figure~\ref{fig:decay} shows the beam intensity and activity curves of the reaction products in $^{108}$Pd activation run.
The solid black line represents the primary proton beam intensity of the ISIS synchrotron.
The orange dashed-dotted, blue dashed, and green dotted lines represent the calculated activity curve ($\lambda n_\mathrm{nucl}(t)$) with $b'=1$ condition for $^{108m}$Pd ($T_{1/2}=6.0~\mathrm{min}$)~\cite{nndc_108}, $^{107g}$Pd ($T_{1/2}=21.7~\mathrm{min}$)~\cite{nndc_107}, and $^{106m}$Pd ($T_{1/2}=131~\mathrm{min}$)~\cite{nndc_106}, respectively.
$P_\mathrm{decay}$ were deduced using Eq.~(\ref{eq:decay}) for each reaction product, as shown in the tables in the next section.

The enrichment of each target was not 100\% and the branching ratio of each reaction product ($b'$) depended on the isotopic composition of the target.
The branching ratios of each reaction product for pure isotopes ($b$) were extracted from $b'$ by solving simultaneous equations of the branching ratio matrix ($B$ and $B'$) and composition matrix ($A$) as follows:
\begin{align}
    A B &= B' \\
    B &= A^{-1}B',\label{eq:matrixcalc}
\end{align}
where 
\begin{align}
    B &= [b_{ij}]\\
    i&:~\mathrm{isotope} \notag \\
    j&:~\mathrm{reaction~products}, \notag
\end{align}
\begin{align}
    B' &= [b'_{tj}]\\
    t&:~\mathrm{enriched~target} \notag \\
    j&:~\mathrm{reaction~products}, \notag
\end{align}
and
\begin{align}
    A &= [a_{tk}]\\
    t&:~\mathrm{enriched~target} \notag \\
    k&:~\mathrm{composition~of~each~isotope}. \notag
\end{align}
The composition matrix ($A$) of the enriched targets is presented in Table~\ref{tab:target_composition}~\cite{Saito2022-ju}.

\section{Results}
\label{sec:result}

In this section, the results of the activation of the five isotopically enriched targets of $^{104,105,106,108,110}$Pd are presented.

\subsection{$^{108}$Pd target}

\begingroup
\squeezetable
\begin{table*}
    \caption{Results of $^{108}$Pd activation.
        Parent nucleus of the decay (Nucleus), spin-parity of the decaying state (State), decay mode (Decay),
        half-life ($T_{1/2}$), decay probability ($P_\mathrm{decay}$),
        $\gamma$-ray energy ($E_\gamma$), $\gamma$-ray intensity ($I_\gamma$),
        the number of emitted $\gamma$ rays ($N_\gamma/(\epsilon_\gamma\epsilon_\mathrm{LT})$), where $\epsilon_\gamma$ is the detection efficiency of the germanium detector and $\epsilon_\mathrm{LT}$ is the analysis livetime, 
        branching ratio deduced by each $\gamma$-ray intensity ($b'_\gamma$)
        and branching ratio for each decaying state ($b'$) are given in the table.
        Decay properties are obtained from ENSDF~\cite{nndc_104,nndc_105,nndc_106,nndc_107,nndc_108}.
        See text for detail.}
    \label{tab:108Pd}
    \begin{ruledtabular}  
    \begin{tabular}{ccccc D{.}{.}{-1} D{.}{.}{-1} D{.}{.}{-1} D{.}{.}{5} D{.}{.}{6} }
        Nucleus & State & Decay &
        $T_{1/2}$ &
        $P_\mathrm{decay} (\%)$ &
        \multicolumn{1}{c}{$E_\gamma$ (keV)} & 
        \multicolumn{1}{c}{$I_\gamma$(\%)\footnotemark[1]} &
        \multicolumn{1}{c}{$N_\gamma/(\epsilon_\gamma \epsilon_\mathrm{LT}) (10^4)$} &
        \multicolumn{1}{c}{$b'_\gamma$(\%)} &
        \multicolumn{1}{c}{$b'$(\%)\footnotemark[2]}\\
        \hline
        $^{108}$Rh & 1$^+$    & $\beta^-$ & 16.8 sec &  99.9 &  434.1\footnotemark[3] & 43.0(30)\footnotemark[4] & 69.1(10) &\\
                   &          &           &          &       &  497.3\footnotemark[3] & 5.2(4)                   & 8.6(7)   &\\
                   &          &           &          &       &  618.9                 & 15.1(13)                 & 20.0(8)  & 11.9(15)\\
                   &&&&&&& \multicolumn{1}{r}{comm.~$\gamma$} & 11.5(9)\\
                   &&&&&& \multicolumn{1}{c}{$\Delta I_\gamma^\mathrm{abs}/I^\mathrm{abs} = 26\%$} & & & 12.(3)\\
                   \cline{2-10}
                   & (5$^+$) & $\beta^-$ & 6.0 min & 97.5 & 404.3  & 26.3(26)               & 4.0(8)   & 1.4(3)\\
                                   & & & & & 434.2\footnotemark[3] & 88.(5)\footnotemark[4] & 69.1(10) & \\
                                   & & & & & 497.4\footnotemark[3] & 19.3(9)                & 8.6(7)   & \\
                                   & & & & & 581.1                 & 60.(4)                 & 6.6(6)   & 1.01(14)\\
                                   & & & & & 614.3                 & 21.0(18)               & 5.9(6)   & 2.6(4)  \\
                                   & & & & & 723.3                 & 10.5(18)               & 4.7(6)   & 4.1(11)  \\
                                   & & & & & 901.3                 & 28.1(26)               & 4.5(7)   & 1.5(3)\\
                                   & & & & & 947.5                 & 49.1(26)               & 8.6(8)   & 1.62(19)\\
                   &&&&&&& \multicolumn{1}{r}{average} & 1.36(10)\\
                   &&&&&&& \multicolumn{1}{r}{comm.~$\gamma$} & 1.37(10)\\
                   &&&&&& \multicolumn{1}{c}{$\Delta I_\gamma^\mathrm{abs}/I_\gamma^\mathrm{abs} = 1.7\%$} & & & 1.37(10)\\
                   \hline
        $^{107}$Rh & 7/2$^+$ & $\beta^-$ & 21.7 min & 89.3  & 277.6 & 1.70(12) & 9.1(7)    & 54.(7) \\
                                                        &&&&& 302.8 & 66.(5)   & 305.3(16) & 46.(5) \\
                                                        &&&&& 312.2 & 4.8(4)   & 24.9(8)   & 52.(6) \\
                                                        &&&&& 321.8 & 2.26(16) & 10.6(8)   & 47.(6) \\
                                                        &&&&& 348.2 & 2.27(16) & 9.2(7)    & 41.(5) \\
                                                        &&&&& 367.3 & 1.91(14) & 7.4(6)    & 39.(5) \\
                                                        &&&&& 392.5 & 8.8(6)   & 28.4(8)   & 44.(4) \\
                                                        &&&&& 567.7 & 1.15(8)  & 4.4(6)    & 39.(6) \\
                                                        &&&&& 670.1 & 2.22(16) & 9.8(6)    & 44.(5) \\
                   &&&&&&& \multicolumn{1}{r}{average} & 44.6(18)\\
                   &&&&&& \multicolumn{1}{c}{$\Delta I_\gamma^\mathrm{abs}/I_\gamma^\mathrm{abs} = 5\%$} & & & 44.6(29)\\
                   \cline{2-10}
                   & 1/2$^-$    & IT        & 0.3--10 sec & 100.0 & 268.4 & 85.3(4)\footnotemark[5] & 194.1(14) & 20.38(20)\\
                   &&&&&&&&& 20.38(20) \\
        \hline
        $^{106}$Rh & 1$^+$ & $\beta^-$ & 30.07 sec & 99.9 & 621.9 & 9.93(12) & 14.2(7) & 12.8(7) \\
                   &&&&&& \multicolumn{1}{c}{$\Delta I_\gamma^\mathrm{abs}/I_\gamma^\mathrm{abs} = 2\%$} & & & 12.8(7)\\
                   \cline{2-10}
                   & (6)$^+$ & $\beta^-$ & 131 min & 60.7  & 221.8  & 6.4(3)   & 2.9(6)  & 6.6(15) & \\
                                                       &&&&& 406.0  & 11.6(7)  & 4.9(7)  & 6.3(10)  & \\
                                                       &&&&& 429.4  & 13.3(21) & 5.2(6)  & 5.8(14) & \\
                                                       &&&&& 450.8  & 24.2(13) & 11.2(7) & 6.8(7)  & \\
                                                       &&&&& 616.1  & 20.2(14) & 10.2(7) & 7.5(9)  & \\
                                                       &&&&& 717.2  & 28.9(15) & 14.1(7) & 7.2(7)  & \\
                                                       &&&&& 748.5  & 19.3(10) & 9.5(8)  & 7.2(8)  & \\
                                                       &&&&& 804.6  & 13.0(11) & 4.3(7)  & 4.9(10)  & \\
                                                       &&&&& 808.4  & 7.4(4)   & 3.5(6)  & 6.9(14) & \\
                                                       &&&&& 825.0  & 13.6(8)  & 4.3(7)  & 4.1(8)  & \\
                                                       &&&&& 1046.7 & 30.4(15) & 13.5(9) & 6.6(6)  & \\
                                                       &&&&& 1200.5 & 11.4(6)  & 5.4(8)  & 7.0(12) & \\
                                                       &&&&& 1224.2 & 8.1(7)   & 3.3(7)  & 5.9(15) & \\
                                                       &&&&& 1529.4 & 17.5(15) & 5.6(8)  & 4.7(9)  & \\
                                                       &&&&& 1573.9 & 6.7(5)   & 1.9(7)  & 4.3(17) & \\
                   &&&&&&& \multicolumn{1}{r}{average} & 6.26(24)\\
                   &&&&&& \multicolumn{1}{c}{$\Delta I_\gamma^\mathrm{abs}/I_\gamma^\mathrm{abs} = 0.8\%$} & & & 6.26(24)\\
                   \hline
        $^{105}$Rh & 7/2$^+$ & $\beta^-$ & 35.3 hour & 34.7\footnotemark[6] & 306.3 & 4.66(5)  & 1.8(4) & 10.2(20)\\
                                                                        &&&&& 319.2 & 16.90(17)& 7.2(5) & 11.0(7)\\
                   &&&&&&& \multicolumn{1}{r}{average} & 10.9(7)\\
                   &&&&&& \multicolumn{1}{c}{$\Delta I_\gamma^\mathrm{abs}/I_\gamma^\mathrm{abs} = 1.8\%$}&&& 10.9(7)\\
                   \cline{2-10}
                   & 1/2$^-$ & IT      & 42.8 sec & 99.8 & 129.8 & 20.2(3)\footnotemark[7] & 8.3(5) & 3.69(24)\\
                   & & & & & & & & & 3.69(24) \\
                   \hline
        $^{104}$Rh & 1$^+$ & $\beta^-$ & 42.3 sec & 98.3 & 555.8 & 2.0(5)\footnotemark[7] & <2.3 & <13.\\
                   \cline{2-10}
                   & 5$^+$ & IT        & 4.34 min & 98.3 & 51.4 & 48.214(5)\footnotemark[7] & 4.2(5) & 0.80(10)\\
                   & & & & & & & & & 0.80(10) \\
        \hline
        $^{107}$Ru & (5/2)$^+$ & $\beta^-$ & 3.75 min & 98.5 & 194.1 & 9.9(17)\footnotemark[7]  & <1.1 & <1.3  \\
        \hline
        $^{105}$Ru & 3/2$^+$ & $\beta^-$ & 4.44 hour & 42.9 & 724.2 & 47.8(6)\footnotemark[7]  & <0.7 & <0.3  \\
        \hline
        $^{104}$Tc & (3$^+$) & $\beta^-$ & 18.3 min & 90.1  & 358.0 & 89.(3)\footnotemark[7]   & <1.9 & <0.22  \\
    \end{tabular}
    \footnotetext[1]{Only the relative uncertainty of the $\gamma$-ray intensity ($\Delta I_\gamma^\mathrm{rel}$) is given in the table, unless noted.}
    \footnotetext[2]{Only the relative uncertainty ($\Delta b'{}^\mathrm{rel}$) is given in the table. For the absolute branching ratio, use $\Delta b'{}^\mathrm{abs}/b' =$ 7\%.}
    \footnotetext[3]{These $\gamma$ rays are observed from the $\beta$ decays of both the ground and isomeric states.}
    \footnotetext[4]{$\Delta I_\gamma^\mathrm{rel}$ of these $\gamma$ rays is not given in the ENSDF database and estimated from other $\Delta I_\gamma$.}
    \footnotetext[5]{$I_\gamma$ of this transition is calculated from 100\% IT decay by considering the conversion coefficient for the E3 multipolarity.}
    \footnotetext[6]{Measured at the RAL offline setup.}
    \footnotetext[7]{Quoted uncertainty includes both $\Delta I_\gamma^\mathrm{rel}$ and $\Delta I_\gamma^\mathrm{abs}$}
    \end{ruledtabular}
\end{table*}
\endgroup

First, we present the results of $^{108}$Pd activation because this data contains most of the essential treatment for data analysis and evaluation of uncertainties.
Table~\ref{tab:108Pd} summarizes the result of $^{108}$Pd activation.
In the activation measurement with the $^{108}$Pd target, the production branching ratios ($b'$) for nine states in $^{108,107,106,105,104}$Rh were obtained.

In this experiment, several $\beta$-delayed $\gamma$ lines were observed in the $\beta$ decay of $^{108g, 108m, 107g, 106m, 105g}$Rh.
The branching ratios were deduced from each $\gamma$-ray intensity ($b'_\gamma$), and the branching ratios of each product ($b'$) were obtained by taking the weighted average.
For an appropriate treatment of the uncertainty, the uncertainty of $I_\gamma$ was divided into two parts: the uncertainty of the relative $\gamma$-ray intensity ($\Delta I_\gamma^\mathrm{rel}$) and that of the absolute intensity ($\Delta I_\gamma^\mathrm{abs}$).
The quoted uncertainty of $b'_\gamma$ in the table includes only $\Delta I_\gamma^\mathrm{rel}$, and $\Delta I_\gamma^\mathrm{abs}$ is added after taking the weighted average.
In the ENSDF database, $I_\gamma^\mathrm{rel}$ is usually given as the relative intensity to the most intense $\gamma$-rays, and the normalization factor for the absolute intensity is written separately in the footnote.
However, $\Delta I_\gamma^\mathrm{rel}$ is occasionally missing for the most intense $\gamma$-rays, for example,  $I_\gamma^\mathrm{rel}=100$ without the quoted uncertainty.
This may be because $\Delta I_\gamma^\mathrm{rel}$ of the most intense $\gamma$ rays propagates to other $\Delta I_\gamma^\mathrm{rel}$.
To set an appropriate weight for the average, the missing $\Delta I_\gamma^\mathrm{rel}$ was estimated from the systematics of $\Delta I_\gamma^\mathrm{rel}$ for the other $\gamma$ rays, assuming that $\Delta I_\gamma^\mathrm{rel}$ was proportional to the square root of $I_\gamma$.

\begin{figure}
    \centering
    \includegraphics[width=8.6cm]{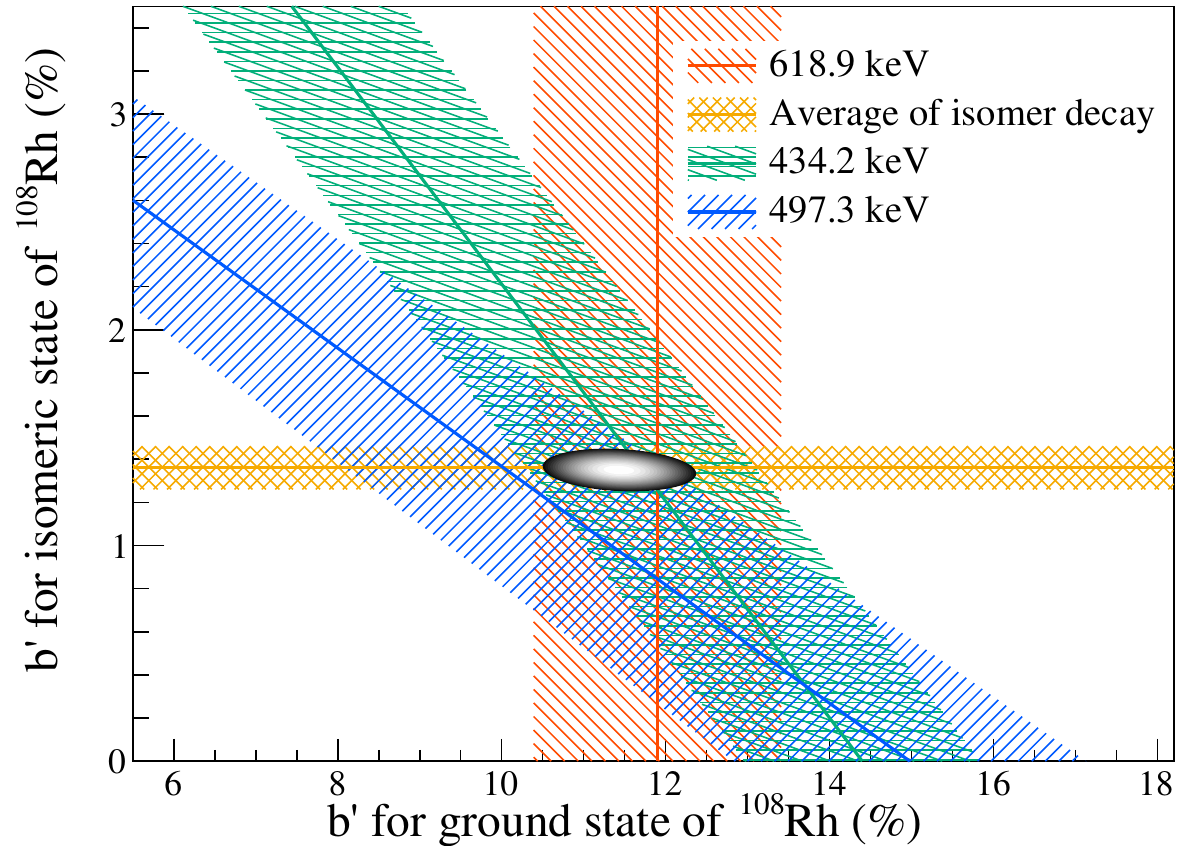}
    \caption{(Color online) Extraction of production branching ratios for the ground and isomeric state in $^{108}$Rh ($b'{}^\textit{gs}$ and $b'{}^\textit{is}$, respectively) including common $\gamma$ rays. 
    Solid lines and hatched area represent $b'$ and its 1$\sigma$ area, respectively, deduced from each $\gamma$-ray intensity.
    The circle at the overlapped area is a 1$\sigma$ uncertainty region of $b'$ for both ground and isomeric states.}
    \label{fig:common}
\end{figure}

There are two $\beta$-decaying states in $^{108}$Rh (0p0n channel): the ground state (1$^+$, $T_{1/2}=16.8$~sec) and the isomeric state ((5$^+$), $T_{1/2}=6.0$~min)~\cite{nndc_108}.
The $\beta$-decay of both states produces excited states in the daughter nucleus of $^{108}$Pd.
Although some $\gamma$ rays are unique for each decay, two $\gamma$ rays at 434.1, and 497.3~keV ($2_1^+ \rightarrow 0_{1}^+$ and $2_2^+ \rightarrow 2_1^+$ transitions in $^{108}$Pd, respectively) are commonly observed in the $\beta$-decay of both states, which are hereafter referred to as common $\gamma$ rays.
$N_\gamma$ of the common $\gamma$ rays contains both ground and isomeric decays; therefore, Eq.~(\ref{eq:branch}) becomes
\begin{equation}
     N_\gamma/(\epsilon_\gamma \epsilon_\mathrm{LT}) = N_\mathrm{cap} ( P_\mathrm{decay}^\textit{gs} b'{}^\textit{gs} I_\gamma^\textit{gs}
     + P_\mathrm{decay}^\textit{is} b'{}^\textit{is} I_\gamma^\textit{is}),
     \label{eq:multigamma}
\end{equation}
where ($P_\mathrm{decay}^\textit{gs}$, $b'{}^\textit{gs}$, $I_\gamma^\textit{gs}$) and ($P_\mathrm{decay}^\textit{is}$, $b'{}^\textit{is}$, $I_\gamma^\textit{is}$) are ($P_\mathrm{decay}$, $b'$, $I_\gamma$) for ground and isomeric states, respectively.
This relation is also used to constrain $b'{}^\textit{gs}$ and $b'{}^\textit{is}$ in addition to $b'_\gamma$ deduced from the unique $\gamma$ rays.
Figure~\ref{fig:common} shows the 1$\sigma$-uncertainty regions constrained by the observed $\gamma$-ray intensities in $b'{}^\textit{gs}$ and $b'{}^\textit{is}$ spaces.
The orange, yellow, green, and blue solid lines and hatches area in the figure represent $b'_\gamma$ and their 1$\sigma$ area as deduced from $\gamma$-ray intensities of 618.9 keV (unique $\gamma$-ray of $^{108g}$Rh decay), average of unique $\gamma$-rays of $^{108m}$Rh decay, 434.2 keV and 497.3 keV, respectively.
$b'{}^\mathit{gs}$ and $b'{}^\mathit{is}$ obtained from the overlapped area were 11.5(9)\% and 1.37(10)\%, respectively, and also expressed in Table~\ref{tab:108Pd} with the label ''comm.~$\gamma$`` (including the intensities of common $\gamma$ rays).

There are also two $\beta$-decaying states in $^{106}$Rh (0p2n channel): the ground state (1$^+$, $T_{1/2}=30.07$~sec) and the isomeric state ((6)$^+$, $T_{1/2}=131$~min).
A common $\gamma$ ray has an energy of 511.85~keV ($2^+\rightarrow0^+$ transition in $^{106}$Pd), which overlaps with the electron annihilation background.
Hence, $b'$ was deduced only from unique $\gamma$ rays at 621.9 keV for $^{106g}$Rh and an average of 15 $\gamma$ lines for $^{106m}$Rh.

\begin{figure}
    \centering
    \includegraphics[width=8.6cm]{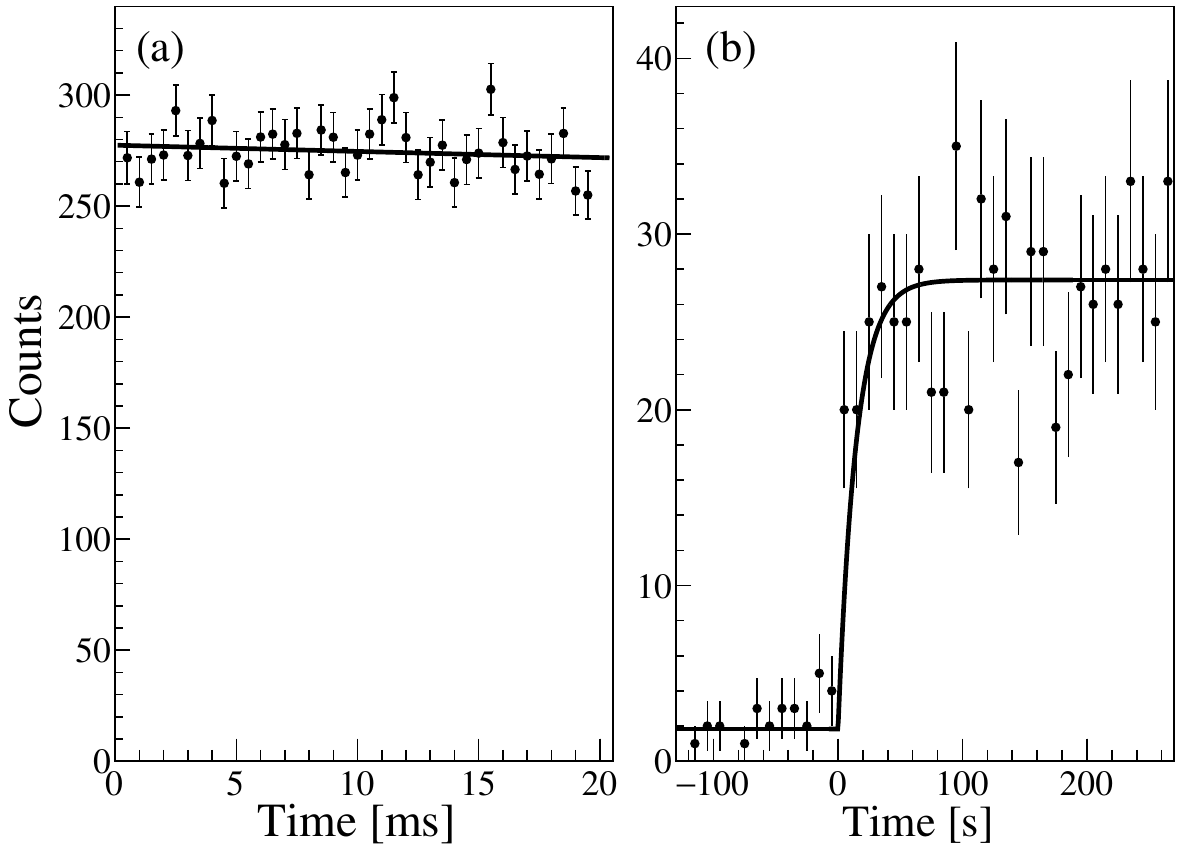}
    \caption{Estimation of the half-life of $^{107m}$Rh.
        (a) 268.4-keV $\gamma$-ray intensity between pulses.
        The time origin of the figure is the timing of the beam arrival.
        (b) 268.4-keV $\gamma$-ray intensity at the beginning of beam irradiation.
        The time origin of the figure is the start timing of beam irradiation.
        }
    \label{fig:pd107m}
\end{figure}

In $^{107}$Rh (0p1n channel), an isomeric state was observed at 268.4~keV (1/2$^-$), that decayed by $\gamma$-ray emission to the ground state (the so-called isomeric transition, IT) with 100\% probability.
Note that $I_\gamma^\textit{is}=0.853(4)$ was not unity because of the electron conversion for this $E3$ transition.
The conversion coefficient was calculated using the \texttt{BrIcc} Conversion Coefficient Calculator~\cite{Kibedi2008-rn}.
Previously, only the lower limit of the half-life of this isomer was known to be $>10$~$\mu$sec~\cite{Kaffrell1986-os}.
If the half-life of the isomeric state is similar to the pulse period of 20~ms, one could observe an exponential decay of the $\gamma$-ray intensity between the beam pulses.
Figure~\ref{fig:pd107m}(a) shows the intensity of 268.4-keV $\gamma$ ray during the interpulse period.
The solid line in the figure represents results fitted with the decay function ($f_\mathrm{decay}(t)$) as follows:
\begin{equation}
    f_\mathrm{decay}(t) = A_1 \exp(-\lambda t),
    \label{eq:decay_curve}
\end{equation}
where $A_1$ is the normalization parameter.
The deduced $\lambda$ by the fitting was consistently zero; hence, only the lower limit of $T_{1/2}>0.3$~sec was obtained.
If the half-life of the isomer is sufficiently long, its half-life can be deduced from the build-up curve at the beginning of beam irradiation, as shown in Fig.~\ref{fig:pd107m}(b).
The solid line in the figure represents the build-up curve ($f_\mathrm{build}(t)$):
\begin{align}
    f_\mathrm{build}(t) = 
    \begin{cases}
    C & (t < 0)\\
    A_2(1-\exp(-\lambda t)) + C & (t \ge 0)
    \end{cases}
    \label{eq:buildup_curve}
\end{align}
where $A_2$ is the normalizaion parameter, $C$ is a constant background term, and $\lambda = \ln(2)/T_{1/2}$ is fixed at $T_{1/2}=10$~sec.
Actually, the present experiment was not designed to measure half-lives using the build-up measurement, which can be performed only at the beginning of beam irradiation and after the incidental beam stops.
Figure~\ref{fig:pd107m}(b) was created by summing three data sets at the beginning of beam irradiation during $^{108}$Pd activation run.
Owing to a lack of statistics, we obtained only the upper limit of the half-life of $^{107m}$Rh with $T_{1/2}<10$~sec.
$P_\mathrm{decay}$ for this isomeric state was deduced from the obtained value $T_{1/2}=$~0.3--10~sec.
Despite the large uncertainty in the half-life, the uncertainty of $P_\mathrm{decay}$ was still negligible ($<$0.1\%).

\begin{figure}
    \centering
    \includegraphics[width=8.6cm]{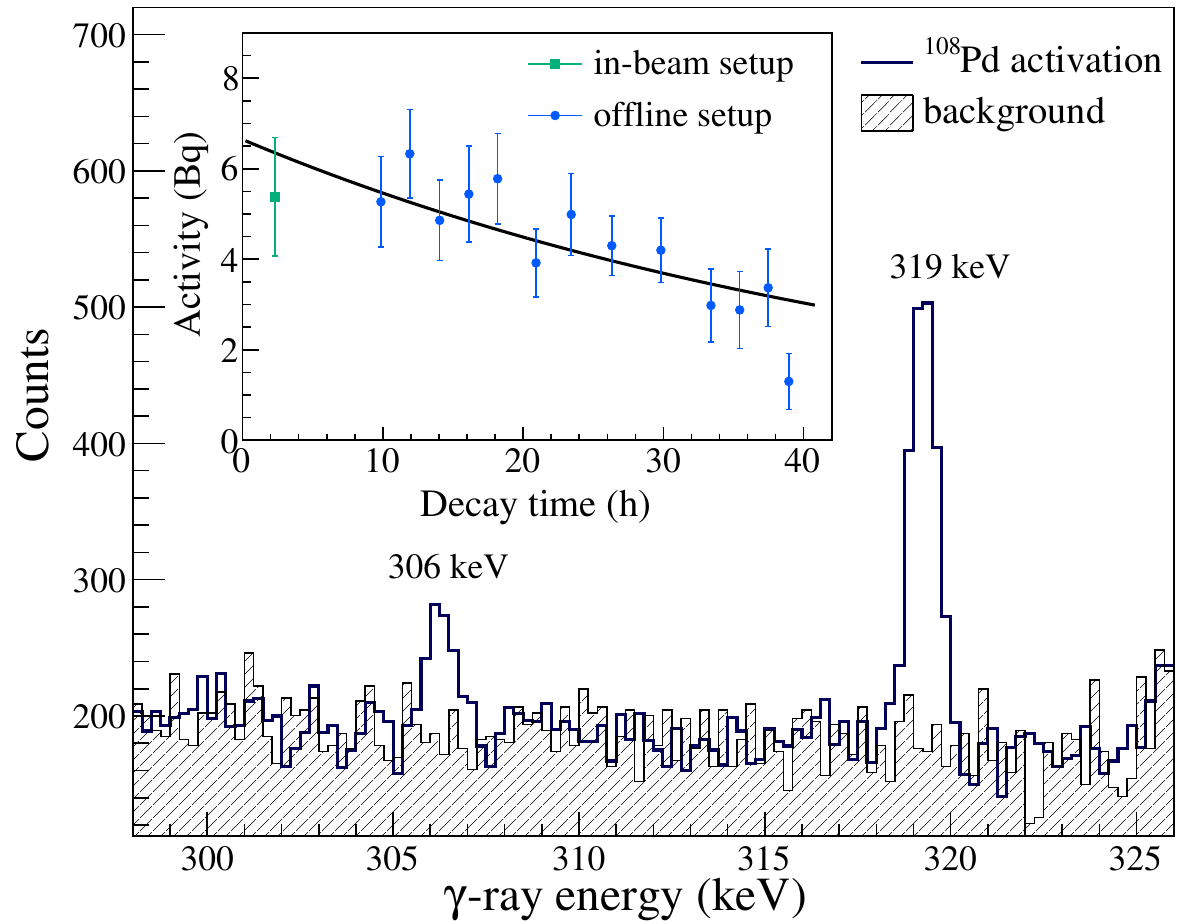}
    \caption{(Color online) $\gamma$-ray spectrum of $^{108}$Pd activation in the RAL offline setup.
    An open histogram represents the $\gamma$-ray spectrum during the offline measurement and a hatched histogram is the backgrounds without targets.
    Two $\gamma$-ray peaks of $^{105g}$Rh decay at 306.3 and 319.2 keV are observed.
    The inset of the figure shows the activity curve of $^{105g}$Rh deduced from the 319.2-keV $\gamma$-ray intensity.
    The first data point is measured by the in-beam setup and others are measured by the RAL offline setup.
    The time origin of the figure (decay time) is the stop time of the muon beam irradiation.
    The solid line represents the decay curve (Eq.~(\ref{eq:decay_curve})) with $T_{1/2}=35.3$~hour.
    }
    \label{fig:decaystation108}
\end{figure}

The $\beta$-delayed $\gamma$ rays of $^{105g}$Rh (0p3n channel, 7/2$^+$, $T_{1/2}=35.3$~hour) were not observed during beam irradiation, and the $\gamma$ intensities were obtained primarily in the offline setup, as shown in Fig.~\ref{fig:decaystation108}.
Two $\gamma$ rays from $^{105g}$Rh decay at 306.3 and 319.2 keV were observed in the spectrum.
The inset of the figure shows the activity of $^{105g}$Rh deduced from the 319.2-keV $\gamma$-ray intensity.
Only the first data point in the inset figure was measured by the in-beam setup immediately after the muon beam stopped; the other points were obtained from the offline setup.
The solid line represents the fit results with the decay curve (Eq.~(\ref{eq:decay_curve})) with a fixed half-life of 35.3~hour.
Although the time integral of Eq.~(\ref{eq:decay_curve}) is equivalent to $N_\gamma/\epsilon_\gamma\epsilon_\mathrm{LT}$, $b'_\gamma$ for both $\gamma$ rays were deduced in the same manner as in-beam activation using Eq.~(\ref{eq:branch}).

In the case of an IT state, one must calculate the differential equation of the decay chain (the so-called Bateman equation~\cite{bateman1910}), namely, Eq.~(\ref{eq:differential}) becomes
\begin{equation}
\begin{split}
    \frac{dn^\mathit{is}(t)}{dt} &= -\lambda^\mathit{is}n^\mathit{is}(t) + b'{}^\mathit{is}n_\mu(t)\\
    \frac{dn^\mathit{gs}(t)}{dt} &= \lambda^\mathit{is}n^\mathit{is}(t) -\lambda^\mathit{gs}n^\mathit{gs}(t) + (b'{}^\mathit{gs}-b'{}^\mathit{is})n_\mu(t),
\end{split}
\label{eq:bateman}
\end{equation}
where $(n^\mathit{gs}, n^\mathit{is})$, $(\lambda^\mathit{gs},\lambda^\mathit{is})$ and $(b'{}^\mathit{gs}, b'{}^\mathit{is})$ are $n_\mathrm{nucl}$, $\lambda$ and $b'$ for ground and isomeric states, respectively.
We defined $b'$ as a cumulative branching ratio, \textit{i.e.} $b'{}^\mathit{gs}$ included both the direct population to the ground state and the population through decay from the other states, including the IT state.
In general, $b'$ of the ground state with the existence of the IT state cannot be deduced from Eq.~(\ref{eq:branch}).
However, the effect of the Bateman equation can be neglected in two cases: $\lambda^\mathit{gs} \ll \lambda^\mathit{is}$ and $P_\mathrm{decay}^\mathit{gs} \simeq P_\mathrm{decay}^\mathit{is} \simeq 1$.
$^{107}$Rh and $^{105}$Rh are the former cases and $^{104}$Rh is the latter.
In $^{108}$Pd activation, $\gamma$-ray decay of $^{104g}$Rh was not observed, and only an upper limit of 13\% for $b'$ of $^{104g}$Rh was obtained.
Practical treatment of the $^{104}$Rh decay data for the $^{104,105,106}$Pd activation runs is explained in the following subsection.

Although $b'$ was not obtained, the detection limits were informative.
We examined the possible products for 0p$x$n ($x=0,1,2,3,4,5$), 1p$x$n ($x=0,1,2$), and 2p2n channels; if no characteristic $\gamma$ ray was observed, the upper limit of $b'$ was deduced whenever possible.
For $^{108}$Pd activation, the upper limits for $^{104g}$Rh (0p4n channel), $^{107}$Ru (1p0n channel), $^{105}$Ru (1p2n channel), and $^{104}$Tc (2p2n channel) were obtained.
The upper limit of $b'$ for $^{106}$Ru (1p1n channel) was not obtained because $^{106}$Ru is a pure $\beta^-$ emitter without $\gamma$-ray emission.
Note that the activation method cannot distinguish between different particle emission processes leading to the same reaction channels; for example, there is no differentiation between one-proton and one-neutron emissions and one-deuteron emission for the production of the 1p1n channel.
However, the 2p2n channel is predominantly produced with alpha emission over the sequential two-proton and two-neutron emissions because of the large binding energy of the alpha particle.

\subsection{$^{106}$Pd target}

\begingroup
\squeezetable
\begin{table*}
    \caption{Results of $^{106}$Pd activation. Same notations as Table~\ref{tab:108Pd}. Decay properties are obtained from ENSDF~\cite{nndc_101,nndc_104,nndc_105,nndc_106,nndc_107}.}
    \label{tab:106Pd}
    \begin{ruledtabular}  
    \begin{tabular}{ccccc D{.}{.}{-1} D{.}{.}{-1} D{.}{.}{-1} D{.}{.}{5} D{.}{.}{6} }
        Nucleus & State & Decay &
        $T_{1/2}$ &
        $P_\mathrm{decay} (\%)$ &
        \multicolumn{1}{c}{$E_\gamma$ (keV)} & 
        \multicolumn{1}{c}{$I_\gamma$(\%)\footnotemark[1]} &
        \multicolumn{1}{c}{$N_\gamma/(\epsilon_\gamma \epsilon_\mathrm{LT}) (10^4)$} &
        \multicolumn{1}{c}{$b'_\gamma$(\%)} &
        \multicolumn{1}{c}{$b'$(\%)\footnotemark[2]}\\
        \hline
        $^{107}$Rh\footnotemark[3] & 7/2$^+$ & $\beta^-$ & 21.7 min   & 99.6  & 302.8 & 66.(5)   & 7.2(10) & 0.37(6)\\
                   &&&&&& \multicolumn{1}{c}{$\Delta I_\gamma^\mathrm{abs}/I_\gamma^\mathrm{abs} = 5.0\%$} &&& 0.37(7)\\
            \cline{2-10}
                   & 1/2$^-$ & IT        & 0.3-10 sec & 100.0 & 268.4 & 85.3(4)\footnotemark[4]  & 3.6(8) & 0.14(3)\\
                   &&&&&&&&& 0.14(3)\\
        \hline
        $^{106}$Rh & 1$^+$   & $\beta^-$ & 30.07 sec   & 100.0 & 616.2\footnotemark[5]  & 0.75(8) & 17.1(8) &\\
                                                           &&&&& 621.9  & 9.93(12)& 45.1(10) & 15.3(4)\\
                                                           &&&&& 1050.4 & 1.56(3) & 7.9(10)  & 17.0(23)\\
                   &&&&&&& \multicolumn{1}{r}{average} & 15.4(4)\\
                   &&&&&&& \multicolumn{1}{r}{comm.~$\gamma$} & 15.3(4)\\
                   &&&&&& \multicolumn{1}{c}{$\Delta I_\gamma^\mathrm{abs}/I_\gamma^\mathrm{abs} = 2\%$} &&& 15.3(5)\\
                   \cline{2-10}
                   & (6)$^+$ & $\beta^-$ & 131 min    & 86.9  & 221.8  & 6.4(3)   & 6.3(8)   & 3.8(6)\\
                                                          &&&&& 406.0  & 11.6(7) & 8.2(8 )   & 2.7(4)\\
                                                          &&&&& 429.4  & 13.3(21) & 8.8(8)   & 2.6(6)\\
                                                          &&&&& 450.8  & 24.2(13) & 17.9(9)  & 2.9(3)\\
                                                          &&&&& 616.1\footnotemark[5]  & 20.2(14) & 17.1(8)  & \\
                                                          &&&&& 717.2  & 28.9(15) & 24.7(9)  & 3.3(3)\\
                                                          &&&&& 748.5  & 19.3(10) & 14.4(8)  & 2.9(3)\\
                                                          &&&&& 793.8  & 5.6(9)   & 5.6(8)   & 3.9(11)\\
                                                          &&&&& 804.6  & 13.0(11) & 9.3(8)   & 2.8(4)\\
                                                          &&&&& 808.4  & 7.4(4)   & 5.9(8)   & 3.1(5)\\
                                                          &&&&& 825.0  & 13.6(8)  & 8.8(8)   & 2.5(3)\\
                                                          &&&&& 1046.7 & 30.4(15) & 20.3(11) & 2.6(2)\\
                                                          &&&&& 1127.7 & 13.7(9)  & 8.9(9)   & 2.5(3)\\
                                                          &&&&& 1200.5 & 11.4(6)  & 9.6(9)   & 3.3(4)\\
                                                          &&&&& 1224.2 & 8.1(7)   & 4.3(8)   & 2.0(5)\\
                                                          &&&&& 1529.4 & 17.5(15) & 10.1(10) & 2.2(4)\\
                                                          &&&&& 1573.9 & 6.7(5)   & 4.6(9)   & 2.7(6)\\
                   &&&&&&& \multicolumn{1}{r}{average} & 2.79(9)\\
                   &&&&&&& \multicolumn{1}{r}{comm.~$\gamma$} & 2.78(9)\\
                   &&&&&& \multicolumn{1}{c}{$\Delta I_\gamma^\mathrm{abs}/I_\gamma^\mathrm{abs} = 1.8\%$} &&& 2.78(9)\\
        \hline
        $^{105}$Rh & 7/2$^+$ & $\beta^-$ & 35.3 hour & 
        18.8\footnotemark[6]      & 306.3\footnotemark[8] & 4.66(5)   & (25.8(12))\footnotemark[8]\\
                              &&&&& 319.2                 & 16.90(17) & 80.1(16)  & 48.3(11)\footnotemark[9] \\
        &&&& 33.1\footnotemark[7] & 306.3\footnotemark[8] & 4.66(5)   & (15.6(6))\footnotemark[8]\\
                              &&&&& 319.2                 & 16.90(17) & 46.7(9)   & 49.6(10)\footnotemark[9] \\
                   &&&&&&& \multicolumn{1}{r}{average} & 49.0(7)\\
                   &&&&&& \multicolumn{1}{c}{$\Delta I_\gamma/I_\gamma = 2.0\%$} &&& 49.0(12)\\
                   \cline{2-10}
                   & 1/2$^-$ & IT & 42.8 sec & 100.0 & 129.8 & 20.2(3)\footnotemark[10] & 105.9(9) & 17.7(4)\\
                   &&&&&&&&& 17.7(4)\\
        \hline
        $^{104}$Rh & 1$^+$ & $\beta^-$ & 42.3 sec & 100.0 & 555.8 & 2.0(5)\footnotemark[10] & 12.9(7) & 22.(6)\\
                   &&&&&&&&& 22.(6) \\
                   \cline{2-10}
                   & 5$^+$ & IT & 4.34 min & 100.0 & 51.4 & 48.214(5)\footnotemark[10] & 5.2(7) & 0.36(5)\\
                   &&&&&&&&& 0.36(5) \\
        \hline
        $^{101}$Rh & 9/2$^+$ & $\epsilon$ & 4.34 day & 
             12.9\footnotemark[6] & 306.9\footnotemark[8] & 81.0(4)\footnotemark[10] & 3.67(13)\footnotemark[8] & 1.2(4)\footnotemark[9]\\
        &&&& 11.7\footnotemark[7] & 306.9\footnotemark[8] & 81.0(4)\footnotemark[10] & 2.7(6)\footnotemark[8]   & 1.0(3)\footnotemark[9]\\
                   &&&&&&& \multicolumn{1}{r}{average} & 1.04(23)\\
                   &&&&&& \multicolumn{1}{c}{$\Delta I_\gamma/I_\gamma = 4.9\%$} &&& 1.04(23)\\
        \hline
        $^{105}$Ru & 3/2$^+$ & $\beta^-$ & 4.44 hour & 65.7 & 469.3 & 18.31(21)\footnotemark[10] & <1.0 & <0.3\\
    \end{tabular}
    \footnotetext[1]{Only the relative uncertainty of the $\gamma$-ray intensity ($\Delta I_\gamma^\mathrm{rel}$) is given in the table unless noted.}
    \footnotetext[2]{Only the relative uncertainty ($\Delta b'{}^\mathrm{rel}$) is given in the table. For the absolute branching ratio, use $\Delta b'{}^\mathrm{abs}/b' =$ 7\%.}
    \footnotetext[3]{Production of $^{107}$Rh in the $^{106}$Pd activation is originates mainly from 0.8\% impurity of $^{108}$Pd in the enriched target.}
    \footnotetext[4]{$I_\gamma$ of this transition is calculated from 100\% IT decay by considering the conversion coefficient for the E3 multipolarity.}
    \footnotetext[5]{These $\gamma$-rays are observed from the $\beta$-decays of both the ground and isomeric states.}
    \footnotetext[6]{Measured at the in-beam setup.}
    \footnotetext[7]{Measured at the RAL offline setup.}
    \footnotetext[8]{306.3-keV $\gamma$-ray from the $^{105g}$Rh decay and 306.9-keV $\gamma$-ray from the $^{101m}$Rh decay are not resolved within the energy resolution of the germanium detector. See the text for a detailed treatment of this $\gamma$-ray intensity.}
    \footnotetext[9]{Quoted uncertainty includes only $\Delta N_\gamma$ and $\Delta I_\gamma$ is added after taking the weighted average.}
    \footnotetext[10]{Quoted uncertainty includes both $\Delta I_\gamma^\mathrm{rel}$ and $\Delta I_\gamma^\mathrm{abs}$}
    \end{ruledtabular}
\end{table*}
\endgroup

Table~\ref{tab:106Pd} summarizes the result of $^{106}$Pd activation.
In the activation measurement with the $^{106}$Pd target, the production branching ratios ($b'$) for the nine states in $^{107,106,105,104,101}$Rh were obtained.

During the activation of the $^{106}$Pd target, decays of $^{107g}$Rh and $^{107m}$Rh were observed in the in-beam spectrum, as shown in Fig.~\ref{fig:gamma}(c) (marked with filled triangles).
The production of $^{107}$Rh originates mainly from the 0.8\% contaminant of $^{108}$Pd in the enriched target.
Note that $b$ of $^{107}$Rh productions from $^{106}$Pd muon capture become zero after calculating Eq.~(\ref{eq:matrixcalc}).

There are two $\beta$-decaying states in $^{106}$Rh (0p0n channel); the ground state (1$^+$, $T_{1/2}=30.07$~sec) and the isomeric state ((6)$^+$, $T_{1/2}=131$~min).
The $\beta$ decays of both states produce excited states of the daughter nucleus of $^{106}$Pd, and 616.2-keV $\gamma$-ray ($2^+_2 \rightarrow 2^+_1$ transition in $^{106}$Pd) is commonly observed from $^{106g}$Rh and $^{106m}$Rh decays.
This common $\gamma$-ray intensity was also used to constrain $b'{}^\mathit{gs}$ and $b'{}^\mathit{is}$ using Eq.~(\ref{eq:multigamma}).

\begin{figure}
    \centering
    \includegraphics[width=8.6cm]{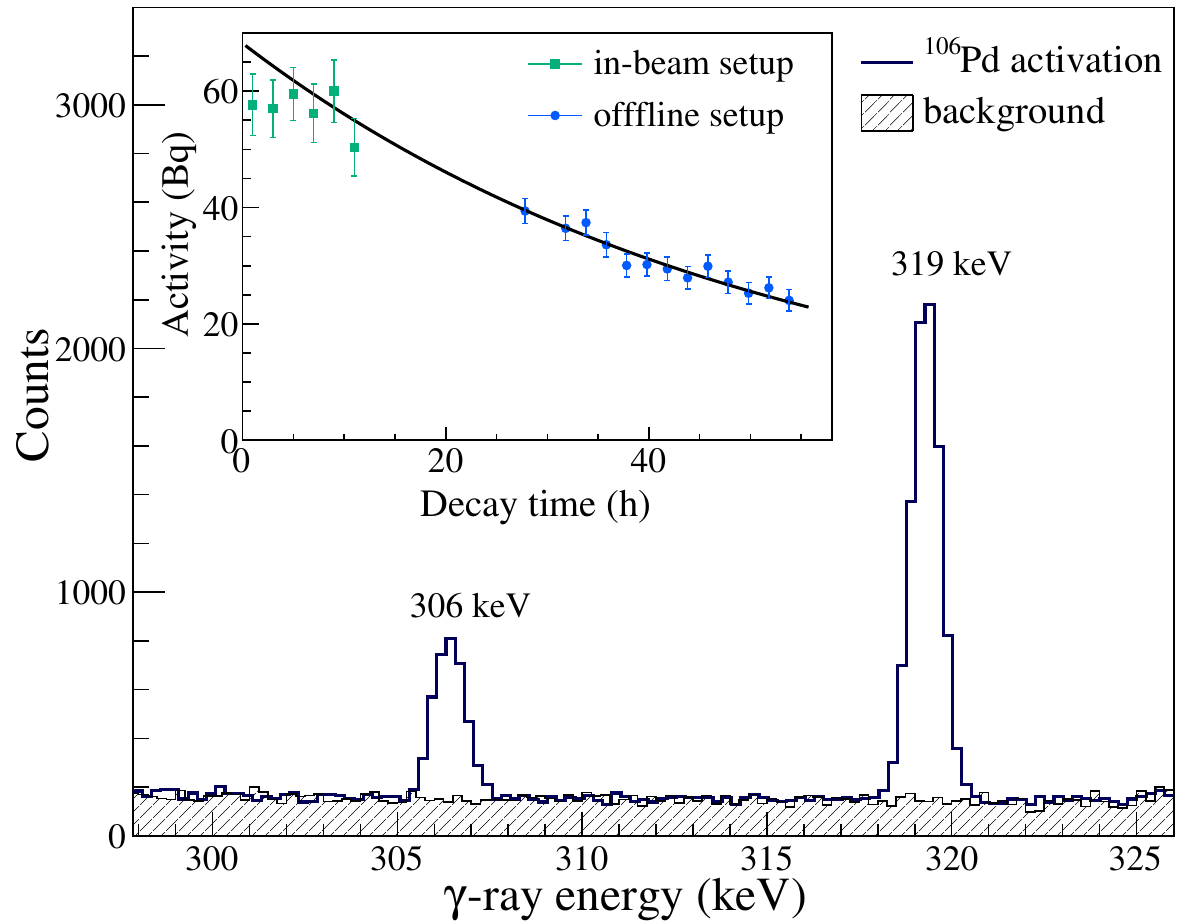}
    \caption{(Color online) $\gamma$-ray spectrum of $^{106}$Pd activation in the RAL offline setup.
    An open histogram represents the $\gamma$-ray spectrum during the offline measurement and a hatched histogram is the background without targets.
    Two $\gamma$-ray peaks at 306 and 319.2 keV are observed.
    The inset of the figure shows an activity curve of $^{105g}$Rh deduced from the 319.2-keV $\gamma$-ray intensity.
    The first six data points are measured by the in-beam setup and others are measured by the RAL offline setup.
    The time origin of the figure (decay time) is the stop time of the muon beam irradiation.
    The solid line represents the decay curve (Eq.~(\ref{eq:decay_curve})) with $T_{1/2}=35.3$~hour.
    }
    \label{fig:decaystation106}
\end{figure}

Figure~\ref{fig:decaystation106} shows the $\gamma$-ray spectrum of the $^{106}$Pd target in the offline setup.
Two peaks at 306 and 319~keV are clearly observed in the spectrum.
While the 319-keV peak corresponds to the decay of $^{105g}$Rh (0p1n channel, 7/2$^+$, $T_{1/2}=35.3$~hour), the 306-keV peak contained both 306.3 keV from $^{105g}$Rh decay and 306.9 keV from $^{101m}$Rh (0p5n channel, 9/2$^+$, $T_{1/2}=4.34$~days).
Because these two $\gamma$-rays were not distinguished within the energy resolution of the germanium detector, $b'$ for $^{105g}$Rh was deduced only from the 319-keV intensity.
The $N_\gamma/(\epsilon_\gamma \epsilon_\mathrm{LT})$ value for 306.3 keV in Table~\ref{tab:106Pd} is the summed intensity of two $\gamma$-rays.
$\gamma$-ray intensity at 306.9 keV of the $^{101m}$Rh decay was extracted by subtraction the 306.3-keV intensity of the $^{105g}$Rh decay estimated by the $I_\gamma$ ratio between 306.3 and 319.2 keV.

$\gamma$-ray decays of $^{105g}$Rh and $^{101}$Rh were observed in both the in-beam and offline setups; $b'$ was obtained from the weighted average of both sets of data.
To treat the uncertainty properly, $\Delta I_\gamma$ (both $\Delta I_\gamma^\mathrm{rel}$ and $\Delta I_\gamma^\mathrm{abs}$) are included in $b'$ after taking the weighted average.

In $^{104}$Rh (0p2n channel), there exists an IT state (5$^+$, $T_{1/2}=4.34$~min) with a longer half-life than that of the ground state (1$^+$, $T_{1/2}=$ 42.3~sec) and the Bateman equation (Eq.~(\ref{eq:bateman})) must be solved to determine $b'$.
The complications in the calculation of the Bateman equation can be avoided by taking a measurement time longer than beam irradiation to achieve the condition $P_\mathrm{decay}^\mathit{gs} \simeq P_\mathrm{decay}^\mathit{is} \simeq 1$, \textit{i.e.} all produced radioactivity decays within the measurement time.
Note that the effective half-life of the ground state at the transient equilibrium is close to that of the isomeric state.
To achieve the above conditions, the measurement time of the $^{106}$Pd activation run in the in-beam setup includes 17.8 hours of beam irradiation time and 1 hour of decay time.
The same conditions were used for $^{104}$Rh measurement in the $^{105}$Pd and $^{104}$Pd activation runs.

In $^{106}$Pd activation, the upper limits for $^{105g}$Ru (1p0n channel) were obtained to be $<0.3$\%.
The upper limits for possible reaction products with $A=102$ isotopes, namely, $^{102g}$Tc (2p2n channel, 1$^+$, $T_{1/2}=5.28$~sec), $^{102m}$Tc (2p2n channel, (4,5), $T_{1/2}=4.35$~min), $^{102g}$Rh (0p4n channel, (1$^-$, 2$^-$), $T_{1/2}=207.3$~days), and $^{102m}$Rh (0p4n channel, 6($^+$), $T_{1/2}=3.74$~year), were not obtained.
$\beta$ decay and electron capture ($\epsilon$) of these four states populated the same $2^+_1$ state in $^{102}$Ru and emitted 475-keV $\gamma$-ray.
Because the 475-keV $\gamma$-ray intensity is the sum of the decay of the four states, no upper limits were deduced for the states involved.

\subsection{$^{104}$Pd target}

\begingroup
\squeezetable
\begin{table*}
    \caption{Results of $^{104}$Pd activation. Same notations as Table~\ref{tab:108Pd}. Decay properties are obtained from ENSDF~\cite{nndc_100,nndc_101,nndc_102,nndc_104}.}
    \label{tab:104Pd}
    \begin{ruledtabular}  
    \begin{tabular}{ccccc D{.}{.}{-1} D{.}{.}{-1} D{.}{.}{-1} D{.}{.}{5} D{.}{.}{6} }
        Nucleus & State & Decay &
        $T_{1/2}$ &
        $P_\mathrm{decay} (\%)$ &
        \multicolumn{1}{c}{$E_\gamma$ (keV)} & 
        \multicolumn{1}{c}{$I_\gamma$(\%)\footnotemark[1]} &
        \multicolumn{1}{c}{$N_\gamma/(\epsilon_\gamma \epsilon_\mathrm{LT}) (10^4)$} &
        \multicolumn{1}{c}{$b'_\gamma$(\%)} &
        \multicolumn{1}{c}{$b'$(\%)\footnotemark[2]}\\
        \hline
        $^{104}$Rh & 1$^+$ & $\beta^-$ & 42.3 sec & 100.0 & 555.8 & 2.0(5) & 5.1(5) & 25.(7)\\
                   &&&&&&&&& 25.(7) \\
                   \cline{2-10}
                   & 5$^+$ & IT & 4.34 min & 100.0 & 51.4 & 48.214(5) & 6.9(4) & 1.41(8)\\
                   &&&&&&&&& 1.41(8) \\
        \hline
        $^{102}$Rh & (1$^-$,2$^-$) & $\epsilon$ & 207.3 day & 2.2\footnotemark[3] & 475.1 & 46.(4) & 0.84(7) & 8.0(12)\\
                   &&&&&&&&& 8.0(12) \\
        \hline
        $^{101}$Rh & 9/2$^+$ & $\epsilon$ & 4.34 day & 20.0\footnotemark[3] & 306.9 & 81.0(4) & 4.59(10) & 2.80(20)\\
                   &&&&&&&&& 2.80(20) \\
        \hline
        $^{103}$Ru & 3/2$^+$ & $\beta^-$ & 39.2 day & 10.1\footnotemark[3] & 497.1 & 91.0(12) & 0.16(5) & 0.18(5)\\
                   &&&&&&&&& 0.18(5) \\
        \hline
        $^{100}$Tc & 1$^+$ & $\beta^-$ & 15.5 sec & 100.0 & 539.5 & 6.60(3) & <1.7 & <2.5\\
    \end{tabular}
    \footnotetext[1]{Quoted uncertainty includes both $\Delta I_\gamma^\mathrm{rel}$ and $\Delta I_\gamma^\mathrm{abs}$}
    \footnotetext[2]{Only the relative uncertainty ($\Delta b'{}^\mathrm{rel}$) is given in the table. For the absolute branching ratio, use $\Delta b'{}^\mathrm{abs}/b' =$ 9\%.}
    \footnotetext[3]{Measured at the UT offline setup.}
    \end{ruledtabular}
\end{table*}
\endgroup

Table~\ref{tab:104Pd} summarizes the results of $^{104}$Pd activation.
In the activation measurement with the $^{104}$Pd target, the production branching ratios ($b'$) for the five states in $^{104,102,101}$Rh and $^{103}$Ru, and the upper limit for $^{100}$Tc were obtained.

Only $^{104g}$Rh (1$^+$, $T_{1/2}=42.3$~sec) and $^{104m}$Rh (5$^+$, $T_{1/2}=4.34$~min) decays (0p0n channel) were observed in the in-beam measurement.
The main product of muon capture of $^{104}$Pd is $^{103}$Rh (0p1n channel), which was not observed in the present experiment because $^{103g}$Rh (1/2$^-$) is a stable isotope.
Although there is an isomeric state of $^{103m}$Rh (7/2$^+$, $T_{1/2}=56.1$~min), the decaying $\gamma$ ray has energy at 39.8 keV, which is below the detection threshold in the present setup.

\begin{figure}
    \centering
    \includegraphics[width=8.6cm]{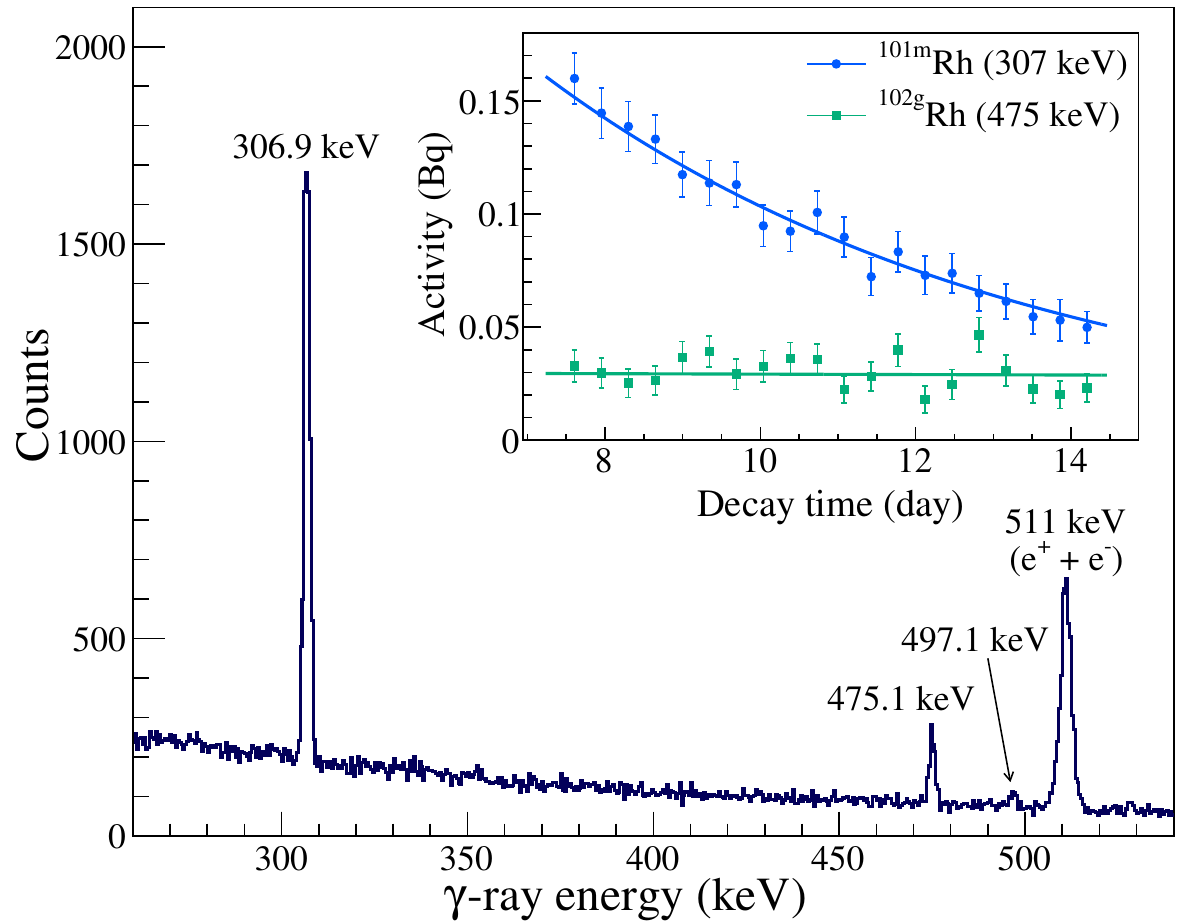}
    \caption{(Color online) $\gamma$-ray spectrum of $^{104}$Pd activation at the UT offline setup.
    Three $\gamma$-ray peaks at 306.9 keV from $^{101m}$Rh decay, at 475.1 keV from $^{102g}$Rh decay, and at 497.1 keV from $^{103g}$Ru decay are observed.
    The inset of the figure shows an activity curve of $^{101m}$Rh and $^{102g}$Rh deduced from the 306.9- and 475.1-keV $\gamma$-ray intensities, respectively.
    The time origin of the figure (decay time) is the stop time of the muon beam irradiation.
    The solid lines represent the decay curve (Eq.~(\ref{eq:decay_curve})) with $T_{1/2}=4.34$~day (blue) and $T_{1/2}=207.3$~day (green) for $^{101m}$Rh and $^{102g}$Rh, respectively.
    }
    \label{fig:decaystation104}
\end{figure}

For the $^{104}$Pd target, the UT offline setup was used for decay measurement for the weak activity of $^{102g}$Rh (0p2n channel, (1$^-$,2$^-$), $T_{1/2}=207.3$~days), $^{101m}$Rh (0p3n channel, 9/2$^+$, $T_{1/2}=4.34$~days), and $^{103g}$Ru (1p0n channel, 3/2$^+$, $T_{1/2}=39.2$~days).
The offline measurement was performed 1 week after beam irradiation.
Figure~\ref{fig:decaystation104} shows the $\gamma$-ray spectrum obtained using the UT offline setup.
Owing to the high sensitivity of the apparatus, three $\gamma$-ray peaks at 306.9 keV from the $^{101m}$Rh decay, 475.1 keV from the $^{102g}$Rh decay, and 497.1 keV from $^{103}$Ru decay were found in the spectrum and $b'$ were deduced.

There are four radioactive states in $A=102$ isotopes that emit the same 475-keV $\gamma$ rays, as explained in the previous subsection.
The two $\beta$-decaying states in $^{102}$Tc (2p0n channel) have short half-lives: $T_{1/2}=5.28$~sec for $^{102g}$Tc and $T_{1/2}=4.35$~min for $^{102m}$Tc; thus, they cannot be measured at the offline setup.
The two electron-capture states in $^{102}$Rh (0p2n channel) have long half-lives: $T_{1/2}=207.3$~days for $^{102g}$Tc and $T_{1/2}=3.74$~year for $^{102m}$Tc, and they cannot be distinguished by the decay curve within a 1-week measurement in the offline setup, as shown in the inset of Fig.~\ref{fig:decaystation104}.
We treated that the observed 475-keV $\gamma$-ray intensity was unique to the $^{102g}$Rh decay and omitted the $^{102m}$Rh decay for the following three reasons.
Because of the difference in $T_{1/2}$, $P_\mathrm{decay}$ for $^{102g}$Rh for the offline measurement was 2.2\%, whereas that for $^{102m}$Rh was 0.35\%.
Hence, the number of decays for $^{102g}$Rh was approximately one order of magnitude higher than that for $^{102m}$Rh.
The population of high-spin isomers by muon capture was systematically smaller than that of the low-spin ground state, as discussed in the next section.
In the decay of $^{102m}$Rh, there are several unique $\gamma$-ray transitions, for example, at 631.3 keV ($I_\gamma=56$\%), 697 keV ($I_\gamma=44$\%), and 766 keV ($I_\gamma=34$\%), in addition to the common $\gamma$ ray at 475.1 keV ($I_\gamma=95$\%).
None of these unique $\gamma$ rays were observed, supporting the exclusion of $^{102m}$Rh decay in the spectrum.

\subsection{$^{105}$Pd target}

\begingroup
\squeezetable
\begin{table*}
    \caption{Results of $^{105}$Pd activation. Same notations as Table~\ref{tab:108Pd}. Decay properties are obtained from ENSDF~\cite{nndc_101,nndc_104,nndc_105}.}
    \label{tab:105Pd}
    \begin{ruledtabular}  
    \begin{tabular}{ccccc D{.}{.}{-1} D{.}{.}{-1} D{.}{.}{-1} D{.}{.}{5} D{.}{.}{6} }
        Nucleus & State & Decay &
        $T_{1/2}$ &
        $P_\mathrm{decay} (\%)$ &
        \multicolumn{1}{c}{$E_\gamma$ (keV)} & 
        \multicolumn{1}{c}{$I_\gamma$(\%)\footnotemark[1]} &
        \multicolumn{1}{c}{$N_\gamma/(\epsilon_\gamma \epsilon_\mathrm{LT}) (10^4)$} &
        \multicolumn{1}{c}{$b'_\gamma$(\%)} &
        \multicolumn{1}{c}{$b'$(\%)\footnotemark[2]}\\
        \hline
        $^{105}$Rh & 7/2$^+$ & $\beta^-$ & 35.3 hour 
        & 18.5\footnotemark[3]    & 306.3\footnotemark[5] & 4.66(5)   & (6.0(9))\footnotemark[5]\\
                              &&&&& 319.2 & 16.90(17) & 12.7(10) & 18.1(15)\footnotemark[6] \\
        &&&& 43.4\footnotemark[4] & 306.3\footnotemark[5] & 4.66(5)   & (18.1(7))\footnotemark[5]\\
                              &&&&& 319.2 & 16.90(17) & 35.4(9) & 21.4(6)\footnotemark[6] \\
                   &&&&&&& \multicolumn{1}{r}{average} & 21.0(5)\\
                   &&&&&& \multicolumn{1}{c}{$\Delta I_\gamma/I_\gamma = 2.0\%$} &&& 21.0(7)\\
                   \cline{2-10}
                   & 1/2$^-$ & IT & 42.8 sec & 100.0 & 129.8 & 20.2(3)\footnotemark[7] & 29.3(7) & 6.45(21)\\
                   &&&&&&&&& 6.45(21)\\
        \hline
        $^{104}$Rh & 1$^+$ & $\beta^-$ & 42.3 sec & 100.0 & 555.8 & 2.0(5)\footnotemark[7] & 18.4(8) & 41.(10)\\
                   &&&&&&&&& 41.(10) \\
                   \cline{2-10}
                   & 5$^+$ & IT & 4.34 min & 100.0 & 51.4 & 48.214(5)\footnotemark[7] & 114.6(13) & 10.55(12)\\
                   &&&&&&&&& 10.55(12) \\
        \hline
        $^{101}$Rh & 9/2$^+$ & $\epsilon$ & 4.34 day & 6.8\footnotemark[3]  & 306.9\footnotemark[5] & 81.0(4)\footnotemark[7] & 2.5(10)\footnotemark[5] & 2.0(7)\footnotemark[6]\\
                                                  &&&& 21.6\footnotemark[4] & 306.9\footnotemark[5] & 81.0(4)\footnotemark[7] & 8.3(8)\footnotemark[5]  & 2.10(21)\footnotemark[6]\\
                   &&&&&&& \multicolumn{1}{r}{average} & 2.10(20)\\
                   &&&&&& \multicolumn{1}{c}{$\Delta I_\gamma/I_\gamma = 4.9\%$} &&& 2.10(23)\\
        \hline
        $^{101}$Tc & 9/2$^+$ & $\beta^-$ & 14.0 min & 100.0 & 306.9\footnotemark[5] & 89.(4)\footnotemark[7] &
        <0.9\footnotemark[8] & <0.05 \\
        \end{tabular}
    \footnotetext[1]{Only the relative uncertainty of the $\gamma$-ray intensity ($\Delta I_\gamma^\mathrm{rel}$) is given in the table, unless noted.}
    \footnotetext[2]{Only the relative uncertainty ($\Delta b'{}^\mathrm{rel}$) is given in the table. For the absolute branching ratio, use $\Delta b'{}^\mathrm{abs}/b' =$ 10\%.}
    \footnotetext[3]{Measured at the in-beam setup.}
    \footnotetext[4]{Measured at the RAL offline setup.}
    \footnotetext[5]{306.3-keV $\gamma$ ray from the $^{105g}$Rh decay and 306.9-keV $\gamma$-ray from the $^{101m}$Rh and possible $^{101}$Tc decays are not resolved within the energy resolution of the germanium detector. See text for a detailed treatment of this $\gamma$-ray intensity.}
    \footnotetext[6]{Quoted uncertainty includes only $\Delta N_\gamma$ and $\Delta I_\gamma$ is added after taking the weighted average.}
    \footnotetext[7]{Quoted uncertainty includes both $\Delta I_\gamma^\mathrm{rel}$ and $\Delta I_\gamma^\mathrm{abs}$}
    \footnotetext[8]{See the text for details regarding the extraction of the upper limit of the $^{101}$Tc production.}
    \end{ruledtabular}
\end{table*}
\endgroup

Table~\ref{tab:105Pd} summarizes the results of $^{105}$Pd activation.
In the activation measurement with the $^{105}$Pd target, the production branching ratios ($b'$) for the five states in $^{105,104,101}$Rh were obtained.

$b'$ for $^{101g}$Rh was extracted, as explained above for the $^{106}$Pd case.
In the case of $^{105}$Pd activation, there could be a production of $^{101g}$Tc (2p2n channel, 9/2$^+$, $T_{1/2}=14.0$~min), which populates the $7/2^+$ state in $^{101}$Ru and emits 306.9-keV $\gamma$-ray.
Thus, the 306.9-keV $\gamma$-ray intensity includes the decays of both $^{101m}$Rh (0p4n channel, 9/2$^+$, $T_{1/2}=4.34$~day) and $^{101g}$Tc.
The decay of $^{101}$Tc can only be observed at the in-beam measurement because of its short half-life.
The consistent $b'$ values for $^{101m}$Rh in the in-beam and offline measurements, as listed in Table~\ref{tab:105Pd}, indicate that $^{101g}$Tc was not observed in the present experiments.
The upper limit of $b'$ for $^{101g}$Tc was extracted from the intensity difference of the 306.9-keV $\gamma$ ray between the in-beam and offline measurements.

\subsection{$^{110}$Pd target}

\begingroup
\squeezetable
\begin{table*}
    \caption{Results of $^{110}$Pd activation. Same notations as Table~\ref{tab:108Pd}. Decay properties are obtained from ENSDF~\cite{nndc_104,nndc_105,nndc_106,nndc_107,nndc_108,nndc_109,nndc_110}.}
    \label{tab:110Pd}
    \begin{ruledtabular}  
    \begin{tabular}{ccccc D{.}{.}{-1} D{.}{.}{-1} D{.}{.}{-1} D{.}{.}{5} D{.}{.}{6} }
        Nucleus & State & Decay &
        $T_{1/2}$ &
        $P_\mathrm{decay} (\%)$ &
        \multicolumn{1}{c}{$E_\gamma$ (keV)} & 
        \multicolumn{1}{c}{$I_\gamma$(\%)\footnotemark[1]} &
        \multicolumn{1}{c}{$N_\gamma/(\epsilon_\gamma \epsilon_\mathrm{LT}) (10^4)$} &
        \multicolumn{1}{c}{$b'_\gamma$(\%)} &
        \multicolumn{1}{c}{$b'$(\%)\footnotemark[2]}\\
        \hline
        $^{110}$Rh & (1$^+$) & $\beta^-$ & 3.35 sec & 99.9 & 357.0 & 1.3(4)  & 2.2(6)  & 13.(7)\\
                                                       &&&&& 373.8\footnotemark[3] & 53.0(5)\footnotemark[4] & 65.1(10)& \\
                                                       &&&&& 439.7\footnotemark[3] & 7.90(27)& 15.1(8) & \\
                                                       &&&&& 796.7 & 4.0(5)  & 3.8(8)  & 7.6(21)\\
                                                       &&&&& 813.7\footnotemark[3] & 2.9(4)  & 4.0(8)  & \\
                   &&&&&&& \multicolumn{1}{r}{average}        & 8.1(20)\\
                   &&&&&&& \multicolumn{1}{r}{comm.~$\gamma$} & 7.3(4)\\
                   &&&&&& \multicolumn{1}{c}{$\Delta I_\gamma^\mathrm{abs}/I^\mathrm{abs} = 38\%$} & & & 7.3(28)\\
                   \cline{2-10}
                   & (6$^+$) & $\beta^-$ & 28.0 sec & 100.0  & 373.8\footnotemark[3] & 89.(4)   & 65.1(10) &\\
                                                           &&&&& 398.6 & 19.8(11) & 4.3(7)   & 1.7(3)\\
                                                           &&&&& 439.8\footnotemark[3] & 29.3(19) & 15.1(8)  &\\
                                                           &&&&& 653.3 & 16.3(14) & 3.0(6)   & 1.5(4)\\
                                                           &&&&& 687.7 & 29.0(21) & 5.2(7)   & 1.43(24)\\
                                                           &&&&& 813.6\footnotemark[3] & 10.2(12) & 4.0(8)   &\\
                                                           &&&&& 838.2 & 21.3(17) & 3.9(8)   & 1.5(3)\\
                                                           &&&&& 904.5 & 17.4(18) & 3.2(8)   & 1.5(4)\\
                   &&&&&&& \multicolumn{1}{r}{average}        & 1.51(14)\\
                   &&&&&&& \multicolumn{1}{r}{comm.~$\gamma$} & 1.55(14)\\
                   &&&&&& \multicolumn{1}{c}{$\Delta I_\gamma^\mathrm{abs}/I^\mathrm{abs} = 4.5\%$} & & & 1.55(15)\\
        \hline
        $^{109}$Rh & 7/2$^+$ & $\beta^-$ & 80.8 sec & 99.7 & 113.4 & 5.7(3)   & 32.4(7)   & 46.(4)    & \\
                                                       &&&&& 178.0 & 7.6(4)   & 49.1(8)   & 52.(4)    & \\
                                                       &&&&& 215.4 & 1.73(11) & 11.6(7)   & 54.(6)    & \\
                                                       &&&&& 245.1 & 1.3(11)  & 7.7(7)    & 48.(7)    & \\
                                                       &&&&& 249.2 & 5.8(3)   & 34.2(8)   & 47.(4)    & \\
                                                       &&&&& 276.3 & 2.16(16) & 14.1(6)   & 52.(6)    & \\
                                                       &&&&& 291.4 & 7.5(4)   & 46.0(11)  & 49.(4)    & \\
                                                       &&&&& 325.3 & 1.46(27) & 9.9(10)   & 54.(15)   & \\
                                                       &&&&& 326.9 & 54.(16)\footnotemark[4]  & 339.8(18) & 50.3(22) & \\
                                                       &&&&& 378.2 & 1.24(11) & 6.2(8)    & 40.(7)    & \\
                                                       &&&&& 426.1 & 7.7(7)   & 53.3(10)  & 55.(7)    & \\
                   &&&&&&& \multicolumn{1}{r}{average} & 49.5(13)\\
                   &&&&&& \multicolumn{1}{c}{$\Delta I_\gamma^\mathrm{abs}/I^\mathrm{abs} = 9.3\%$} & & & 50.(5)\\
        \hline
        $^{108}$Rh & 1$^+$    & $\beta^-$ & 16.8 sec &  99.9 &  434.1\footnotemark[3]  & 43.(4)\footnotemark[4]& 95.7(12)& \\
                   &          &           &           &       &  497.3\footnotemark[3] &  5.2(4)               & 15.3(9) & \\
                   &          &           &           &       &  618.9                 & 15.1(13)              & 20.4(14)& 10.8(15)\\
                   &          &           &           &       &  931.7\footnotemark[3] & 1.25(13)              & 4.8(9)  & \\
                   &&&&&&& \multicolumn{1}{r}{comm.~$\gamma$} & 10.1(10)\\
                   &&&&&& \multicolumn{1}{c}{$\Delta I_\gamma^\mathrm{abs}/I^\mathrm{abs} = 26\%$} & & & 10.1(28)\\
                   \cline{2-10}
                   & (5$^+$) & $\beta^-$ & 6.0 min & 98.5  & 404.3                 & 26.3(26)               & 11.0(8)  & 3.4(5)\\
                                                       &&&&& 434.2\footnotemark[3] & 88.(5)\footnotemark[4] & 95.7(12) & \\
                                                       &&&&& 497.4\footnotemark[3] & 19.3(9)                & 15.3(9)  & \\
                                                       &&&&& 581.1                 & 60.(4)                 & 26.5(24) & 3.4(5)\\
                                                       &&&&& 614.3                 & 21.0(18)               & 13.1(7)  & 5.0(7)\\
                                                       &&&&& 723.3                 & 10.5(18)               & 4.0(7)   & 3.1(9)\\
                                                       &&&&& 901.3                 & 28.1(26)               & 13.7(9)  & 4.0(6)\\
                                                       &&&&& 931.7\footnotemark[3] & 12.3(18)               & 4.8(9)   & \\
                                                       &&&&& 947.5                 & 49.1(26)               & 24.1(10) & 4.0(3)\\
                                                       &&&&& 1234.3                & 8.8(18)                & 4.3(9)   & 4.0(14)\\
                   &&&&&&& \multicolumn{1}{r}{average} & 3.87(21)\\
                   &&&&&&& \multicolumn{1}{r}{comm.~$\gamma$} & 3.76(20)\\
                   &&&&&& \multicolumn{1}{c}{$\Delta I_\gamma^\mathrm{abs}/I_\gamma^\mathrm{abs} = 1.7\%$} &&& 3.76(21)\\
                   \hline
        $^{107}$Rh & 7/2$^+$ & $\beta^-$ & 21.7 min & 94.6  & 302.8 & 66.(5)   & 73.2(12) & 9.3(10)  \\
                                                        &&&&& 312.2 & 4.8(4)   & 6.8(7)   & 12.0(19)\\
                                                        &&&&& 321.8 & 2.26(16) & 2.7(8)   & 10.0(31) \\
                                                        &&&&& 348.2 & 2.27(16) & 3.1(7)   & 11.7(29) \\
                                                        &&&&& 392.5 & 8.8(6)   & 9.2(7)   & 8.8(11)  \\
                   &&&&&&& \multicolumn{1}{r}{average} & 9.6(7)\\
                   &&&&&& \multicolumn{1}{c}{$\Delta I_\gamma^\mathrm{abs}/I_\gamma^\mathrm{abs} = 5\%$} &&& 9.6(8)\\
                   \cline{2-10}
                   & 1/2$^-$    & IT        & 0.3--10 sec & 100.0 & 268.4 & 85.3(4)\footnotemark[5] & 35.1(8) & 3.28(8) \\
                   &&&&&&&&& 3.28(8) \\
        \hline
        $^{106}$Rh & 1$^+$ & $\beta^-$ & 30.07 sec & 99.9 & 621.9 & 9.93(12) & 3.7(8) & 3.0(6) \\
                   &&&&&& \multicolumn{1}{c}{$\Delta I_\gamma^\mathrm{abs}/I_\gamma^\mathrm{abs} = 2\%$} &&& 3.0(6)\\
                   \cline{2-10}
                   & (6)$^+$ & $\beta^-$ & 131 min & 69.7  & 406.0  & 11.6(7)  & 2.4(7) & 2.4(7) \\
                                                       &&&&& 450.8  & 24.2(13) & 3.3(7) & 1.6(3) \\
                                                       &&&&& 717.2  & 28.9(15) & 3.0(7) & 1.2(3) \\
                                                       &&&&& 748.5  & 19.3(10) & 4.2(7) & 2.5(5) \\
                   &&&&&&& \multicolumn{1}{r}{average} & 1.61(19)\\
                   &&&&&& \multicolumn{1}{c}{$\Delta I_\gamma^\mathrm{abs}/I_\gamma^\mathrm{abs} = 0.8\%$} &&& 1.61(19)\\
        \hline
        $^{105}$Rh & 1/2$^-$ & IT & 42.8 sec & 99.8 & 129.8 & 20.2(3)\footnotemark[6] & 2.2(5) & 0.85(22)\\
                   &&&&&&&&& 0.85(22) \\
        \hline
        $^{104}$Rh & 1$^+$ & $\beta^-$ & 42.3 sec & 98.9 & 555.8 & 2.0(5)\footnotemark[6]    & <1.4 & <7.  \\
                   \cline{2-10}
                   & 5$^+$ & IT        & 4.34 min & 98.9 & 51.4  & 48.214(5)\footnotemark[6] & <0.9 & <0.15 \\
        \hline
        $^{109}$Ru & (5/2$^+$) & $\beta^-$ & 34.4 sec & 99.9 & 206.3 & 20.7(15)\footnotemark[6] & <0.5 & <0.23  \\
        \hline
        $^{108}$Ru & 0$^+$     & $\beta^-$ & 4.55 min & 98.9 & 164.9 & 28.0(8)\footnotemark[6]  & <1.1 & <0.3  \\
        \hline
        $^{107}$Ru & (5/2)$^+$ & $\beta^-$ & 3.75 min & 99.1 & 194.1 & 9.9(17)\footnotemark[6]  & <0.4 & <0.4  \\
        \hline
        $^{106}$Tc & (2$^+$)   & $\beta^-$ & 35.6 sec & 99.9 & 270.1 & 55.8(17)\footnotemark[6] & <1.7 & <0.26 \\
        
    \end{tabular}
    \footnotetext[1]{Only the relative uncertainty of the $\gamma$-ray intensity ($\Delta I_\gamma^\mathrm{rel}$) is given in the table unless noted.}
    \footnotetext[2]{Only the relative uncertainty ($\Delta b'{}^\mathrm{rel}$) is given in the table. For the absolute branching ratio, use $\Delta b'{}^\mathrm{abs}/b' =$ 9\%.}
    \footnotetext[3]{These $\gamma$-rays are observed from the $\beta$-decays of both the ground and isomeric states.}
    \footnotetext[4]{$\Delta I_\gamma^\mathrm{rel}$ of these $\gamma$-rays is not given in the ENSDF database and estimated from other $\Delta I_\gamma$.}
    \footnotetext[5]{$I_\gamma$ of this transition is calculated from 100\% IT decay by considering the conversion coefficient for the E3 multipolarity.}
    \footnotetext[6]{Quoted uncertainty includes both $\Delta I_\gamma^\mathrm{rel}$ and $\Delta I_\gamma^\mathrm{abs}$.}
    \end{ruledtabular}
\end{table*}
\endgroup

Table~\ref{tab:110Pd} summarizes the results of $^{110}$Pd activation.
In the activation measurement with the $^{110}$Pd target, the production branching ratios ($b'$) for the 10 states in $^{110,109,108,107,106,105}$Rh were obtained.
For this target, only in-beam measurement was conducted.

There are two $\beta$-decaying states in $^{110}$Rh (0p0n channel): the ground state ((1$^+$), $T_{1/2}=3.35$~sec) and the isomeric state ((6$^+$), $T_{1/2}=28.0$~sec).
The $\beta$ decays of both states produce excited states of the daughter nucleus of $^{110}$Pd, and three $\gamma$-rays at 373.8, 439.7, and 813.6 keV ($2^+_1 \rightarrow 0^+_0$, $2^+_2 \rightarrow 0^+_0$, and $2^+_2 \rightarrow 2^+_1$ transitions in $^{110}$Pd) were commonly observed from $^{110g}$Rh and $^{110m}$Rh decays.
These common $\gamma$-ray intensities were also used to constrain $b'{}^\mathit{gs}$ and $b'{}^\mathit{is}$ using Eq.~(\ref{eq:multigamma}).
The same treatment was applied to the commonly observed $\gamma$ rays in $^{108}$Rh decays (0p2n channel).

In $^{107}$Rh (0p3n channel), an isomeric state at 268.4~keV (1/2$^-$) was observed.
The half-life of this state was obtained from $^{108}$Pd activation data, as shown in Fig.~\ref{fig:pd107m}.
Because the statistics for the $^{107m}$Rh decay in the $^{110}$Pd data were lower than those in the $^{108}$Pd data, the same half-life value of $T_{1/2}=$ 0.3--10~sec was used in the analysis.

\subsection{Branching ratios for each isotope}
\label{sec:branch}

\begingroup
\begin{table*}
    \caption{Transposed matrix of the production branching ratio ($B^\top$), absolute uncertainty ($\Delta b^\mathrm{abs}/b$), and total yields ($\sum b^\mathrm{tot}$) for $^{104,105,106,108,110}$Pd.
    Quoted uncertainty on each $b$ is a relative uncertainty ($\Delta b^\mathrm{rel}$) and the absolute uncertainty ($\Delta b^\mathrm{abs}$) is separately written at the bottom.
    Quoted uncertainty on the total yields is the sum of relative and absolute uncertainties.
    }
    \label{tab:result}
    \begin{ruledtabular}  
    \begin{tabular}{cccc D{.}{.}{-1} D{.}{.}{-1} D{.}{.}{-1} D{.}{.}{-1} D{.}{.}{-1} }
        \multicolumn{4}{c}{Reaction products} & \multicolumn{5}{c}{Branching ratio for each isotope ($b$) (\%)} \\
        Nucleus & State & Decay\footnotemark[1] & $T_{1/2}$
        & \multicolumn{1}{c}{$^{104}$Pd} 
        & \multicolumn{1}{c}{$^{105}$Pd} 
        & \multicolumn{1}{c}{$^{106}$Pd} 
        & \multicolumn{1}{c}{$^{108}$Pd} 
        & \multicolumn{1}{c}{$^{110}$Pd} \\
        \hline
        $^{110}$Rh & (1$^+$)       & $\beta^-$  & 3.35 sec     &           &            &           &            & 7.4(28) \\
                   & (6$^+$)       & $\beta^-$  & 28.0 sec     &           &            &           &            & 1.57(16)\\
        $^{109}$Rh & 7/2$^+$       & $\beta^-$  & 80.8 sec     &           &            &           &            & 50.(5)  \\
        $^{108}$Rh & 1$^+$         & $\beta^-$  & 16.8 sec     &           &            &           &  12.(3)    & 10.2(28)\\
                   & (5$^+$)       & $\beta^-$  & 6.0 min      &           &            &           &  1.45(10)  & 3.82(21)\\
        $^{107}$Rh & 7/2$^+$       & $\beta^-$  & 21.7 min     &           &            &           &  48.(3)    & 9.4(8)  \\
                   & 1/2$^-$       & IT         & 0.3--10.0 sec&           &            &           &  21.72(22) & 3.19(8) \\
        $^{106}$Rh & 1$^+$         & $\beta^-$  & 30.07 sec    &           &            &  15.5(5)  &  13.5(7)   & 2.9(6)  \\
                   & (6)$^+$       & $\beta^-$  & 131 min      &           &            &  2.77(9)  &  6.65(26)  & 1.58(20)\\
        $^{105}$Rh & 7/2$^+$       & $\beta^-$  & 35.3 hour    &           &  20.7(7)   &  49.5(13) &  11.2(8)   &         \\
                   & 1/2$^-$       & IT         & 42.8 sec     &           &  6.32(21)  &  17.9(4)  &  3.75(25)  & 0.73(22)\\
        $^{104}$Rh & 1$^+$         & $\beta^-$  & 42.3 sec     &  25.(7)   &  41.(11)   &  22.(6)   &  <12.      & <7.     \\
                   & 5$^+$         & IT         & 4.34 min     &  1.31(8)  &  10.77(12) &  0.29(5)  &  0.76(11)  & <0.11   \\
        $^{102}$Rh & (1$^-$,2$^-$) & $\epsilon$ & 207.3 day    &  8.2(12)  &            &           &            &         \\
        $^{101}$Rh & 9/2$^+$\footnotemark[2] &$\epsilon$ & 4.34 day&  2.82(21) & 2.12(23) & 1.04(23)  &            &       \\
        $^{103}$Ru & 3/2$^+$       & $\beta^-$  & 39.2 day     &  0.18(6)  &            &           &            &       \\
        \hline
        \multicolumn{4}{c}{Absolute uncertainty ($\Delta b^\mathrm{abs}/b$)}
        & \multicolumn{1}{c}{9\%}
        & \multicolumn{1}{c}{10\%}
        & \multicolumn{1}{c}{7\%}
        & \multicolumn{1}{c}{7\%}
        & \multicolumn{1}{c}{9\%}\\
        \hline
        \multicolumn{4}{c}{Total yield ($\sum b^\mathrm{tot}$)}
        & \multicolumn{1}{c}{36(6)\%}
        & \multicolumn{1}{c}{64(13)\%}
        & \multicolumn{1}{c}{91(12)\%}
        & \multicolumn{1}{c}{93(11)\%}
        & \multicolumn{1}{c}{88(14)\%}\\
    \end{tabular}
    \footnotetext[1]{Only the observed decay mode is given in the table.}
    \footnotetext[2]{This is an isomeric state. The decay of the $^{101}$Rh ground state ($1/2^-$, $T_{1/2} =$~3.3 years) is not observed in the present experiment.}
    \end{ruledtabular}
\end{table*}
\endgroup

Table~\ref{tab:result} lists the transposed matrix of the production branching ratios ($B^\top$) for $^{104,105,106,108,110}$Pd calculated using Eq.~(\ref{eq:matrixcalc}).
The production branching ratios for charged particle emission channels are summarized in Table~\ref{tab:chargedparticle}.

There are two uncertainties in $b$: relative ($\Delta b^\mathrm{rel}$) and absolute ($\Delta b^\mathrm{abs}$).
$\Delta b^\mathrm{rel}$ is dominated by the statistical uncertainty of the $\gamma$-ray intensity ($\Delta N_\gamma$) and the total uncertainty of the $\gamma$-ray intensity per decay of the reaction products ($\Delta I_\gamma$).
Both the relative and absolute uncertainties of $I_\gamma$ ($\Delta I_\gamma^\mathrm{rel}$ and $\Delta I_\gamma^\mathrm{abs}$, respectively) reflect the relative uncertainty of $b$ ($\Delta b^\mathrm{rel}$). 
In most cases, $\Delta b^\mathrm{rel}$ was dominated by the $\Delta I_\gamma$.
The uncertainty in $P_\mathrm{decay}$, which is propagated from $\Delta T_{1/2}$, was negligible.
The uncertainty in the absolute branching ratio originates from $\Delta N_\mathrm{cap}/N_\mathrm{cap}=2$\%, $\Delta \epsilon_\mathrm{stop}/\epsilon_\mathrm{stop}=$ 1--4\% depending on the targets, $\Delta P_\mathrm{cap}/P_\mathrm{cap}=$ 1\%, and $\Delta \epsilon_\gamma/\epsilon_\gamma=$ 3\%.
Only $\Delta b^\mathrm{rel}$ is listed for each $b$ in Table~\ref{tab:result} and $\Delta b^\mathrm{abs}/b$ is given separately at the bottom of the table.

\section{Discussion}
\label{sec:discussion}

\subsection{Production branching ratio of muon capture}

\begin{figure}
    \centering
    \includegraphics[width=8.6cm]{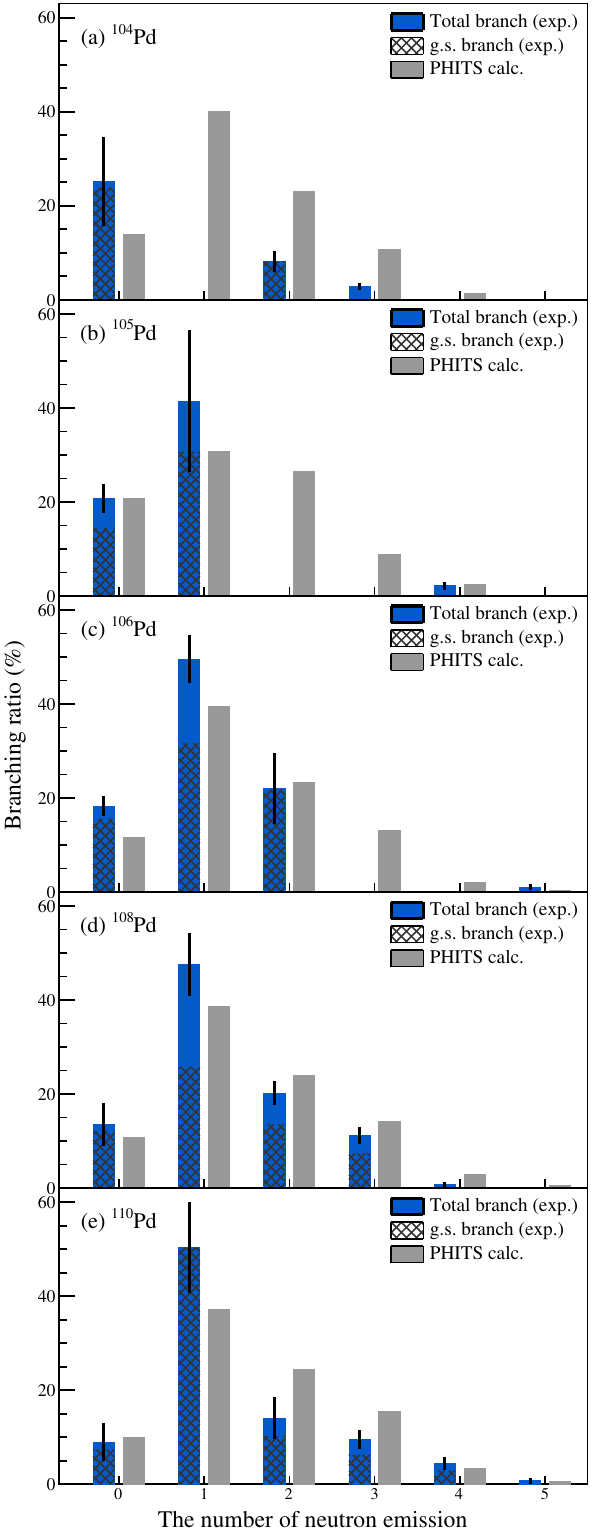}
    \caption{(Color online) Measured branching ratios for each isotope produced by muon capture in the present study and the calculated branching ratios using the particle and heavy ion transport system (PHITS) code.}
    \label{fig:nmulti}
\end{figure}

\begin{figure}
    \centering
    \includegraphics[width=8.6cm]{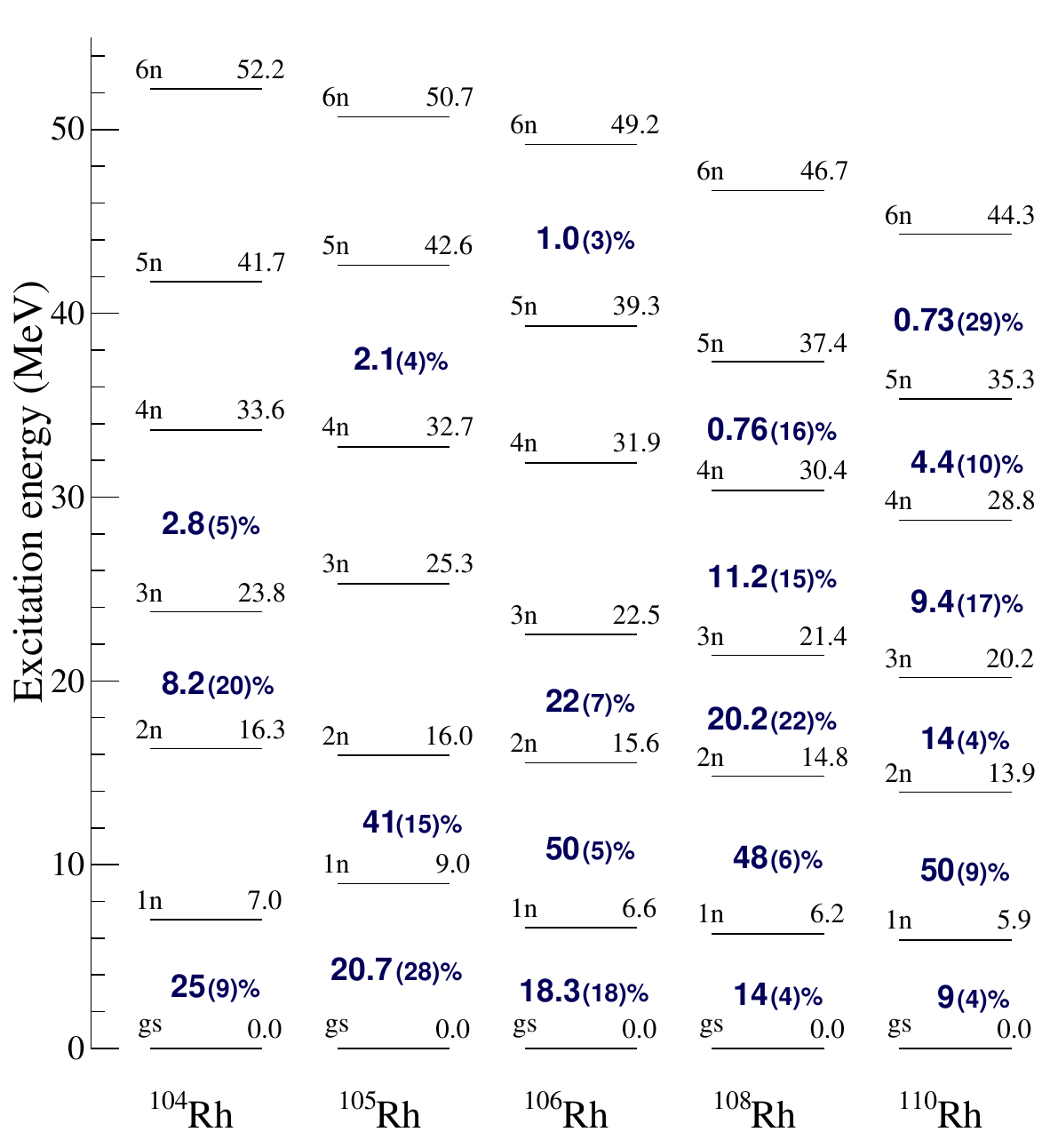}
    \caption{(Color online) Neutron emission thresholds of the rhodium isotopes.
    The measured total branching ratios ($b^\mathrm{tot}$) are shown between the threshold levels assuming that excitation above the threshold energy decays via neutron evaporation with the multiplicity of the level just below.}
    \label{fig:threshold}
\end{figure}

Figure~\ref{fig:nmulti} presents the total branching ratios following the muon capture reaction for the five palladium isotopes obtained in this study.
If the reaction residue has an IT state, the total branching ratio ($b^\mathrm{tot}$) is equal to that of the ground state ($b^\mathit{gs}$), and, if the residue has two $\beta$-decaying states (ground and isomeric states), the total branching ratio is the sum of $b^\mathit{gs}$ and $b^\mathit{is}$.
The total yield of the produced nuclei ($\sum b^\mathrm{tot}$) is listed in Table~\ref{tab:result}.
Production yields of approximately 90\% were obtained for $^{106,108,110}$Pd, whereas only a part of the production yield was measured for $^{104,105}$Pd, primarily because of the lack of the $b$ value of the stable $^{103}$Rh.
The total branching ratios of muon capture were approximately 10--20\% for the 0n channel, 50\% for the 1n channel, 10--20\% for the 2n channel, and the rest for the other channels.
Although the general trend of this neutron multiplicity distribution was previously indicated~\cite{Macdonald1965-tw}, the results of the present study provide the first concrete experimental data for the distribution of the production branching ratios without any theoretical estimation or assumptions in the interpretation of the data analysis.

To compare the obtained production branching ratios in the present study with those of a model calculation, the total branching ratios were calculated with the Monte Carlo simulation using the particle and heavy ions transport code system (PHITS)~\cite{Sato2013-gq}, as shown in Fig.~\ref{fig:nmulti}.
Muon interaction models have recently been implemented in the PHITS code~\cite{Abe2017-xw}.
In this model, the neutron energy produced by muon capture (Eq.~(\ref{eq:capture_mup})) was sampled from the excitation function proposed by Singer~\cite{Singer1962-ww}, in which the momentum distribution of the proton inside the nucleus was estimated using the model proposed by Amado~\cite{Amado1976-cy}.
The time evolution of the initial neutron energy to the compound nucleus was calculated using Jaeri Quantum Molecular Dynamics (JQMD)~\cite{Niita1995-jqmd, Ogawa2015-gu} and the sequential evaporation process was calculated using Generalized Evaporation Model (GEM)~\cite{Furihata2000-nl}.
The model calculation reproduced the general trend of the obtained branching ratios rather well.

The neutron emission thresholds of the compound nuclei (rhodium isotopes) are shown in Fig.~\ref{fig:threshold}.
Threshold energies were calculated using the mass table of NUBASE2016~\cite{Audi2017-ie}.
The measured $b^\mathrm{tot}$ values are shown in the figure between the threshold levels, assuming that the excitation above the threshold energy decays via neutron emissions with the neutron multiplicity of the level just below.

Approximately 50\% of muon capture produces a 1n channel of the residual nucleus.
Because the typical energy for one neutron emission is approximately 7 MeV and that for two neutron emissions is approximately 15 MeV, the center of the excitation energy distribution by muon capture is suggested to be approximately 10 MeV.
First, muon capture is supposed to excite similar bound levels as the $(n,p)$ charge exchange reaction.
Therefore, the excited states populated by muon capture follow the Gamow-Teller (GT) strength observed in the $(n,p)$ reaction, and GT 1$^+$ transitions are important (but not the only transitions)~\cite{Measday2001-hi}.
Although $(n,p)$ reaction studies have not been performed on palladium isotopes, the monopole and dipole strength for heavy nuclei in the $^{120}\mathrm{Sn}(n,p)$, $^{181}\mathrm{Ta}(n,p)$, and $^{238}\mathrm{U}(n,p)$ reactions showed the largest cross section at around 10--15 MeV~\cite{Alford2002-mu}.
The highest production branching ratios for the 1n channel indicate the importance of the GT strength in muon capture.
Second, evaporation neutrons are not the only mechanism of the decay process of muon capture, and the production yield of the 1n residue includes high-energy single-neutron emission from the direct and preequilibrium processes.
Singer introduced the concept of surface effects in muon capture, which increases the single-neutron production and improves the agreement of its production probability~\cite{Singer1962-ww}.
The neutron energy spectrum indicates that the portion of the direct and preequilibrium processes is approximately 15\% for heavy nuclei~\cite{Schroder1974-jb}, and is also similar to that of palladium isotopes~\cite{Saito2022-su}.
In the PHITS calculation, the direct and preequilibrium processes are implemented in JQMD.
In this model, the energetic neutron produced by muon capture causes cascade scattering with nucleons in the nucleus, and the outgoing neutron in the scattering process represents the direct or preequilibrium processes.
However, the PHITS calculation underestimates these effects by approximately 5\%, whereas experimental observations indicate that the effects are greater than 10\%~\cite{Schroder1974-jb, Saito2022-su}.
The underestimation of $b$ for the 1n channel by the PHITS calculation may be due to the small direct and preequilibrium components in the model.

There is a clear isotope dependence on the branching ratio for the 0n channel.
$b^\mathrm{tot}$ for the 0n channel increased as the target mass number decreased.
This trend can be interpreted as following two reasons: (1) higher neutron emission thresholds for neutron-deficient nuclei and/or (2) the low excitation energy of the compound states populated by muon capture for proton-rich nuclei.
In the PHITS calculation, namely, in GEM, the threshold effect is implemented from the mass table of NUBASE2016 and predicts a gentle increase in the branching ratio as the neutron emission threshold increases.
However, our results showed a more drastic increase in the production of the 0n channel.
The PHITS calculation also overestimated the production of 2n and 3n residues for muon capture of all palladium isotopes.
Hence, the model in PHITS, namely, the Singer model, may overestimate the excitation energy produced by muon capture.

The significant population of the high-spin isomeric state in the odd-odd rhodium isotopes, namely, $^{104m}$Rh (5$^+$), $^{106m}$Rh (6$^+$), and $^{108m}$Rh (5$^+$), helps understand the origin of the angular momentum introduced into the compound nucleus by muon capture.
Because the initial angular momenta of muon capture are the spin of the muon (1/2), zero for the orbital angular momentum of the muonic atom (1s state), and the orbital angular momentum of the captured proton in the nuclear medium, the spin state of the compound state is supposed to have low spin.
The recoil of the emitted neutrino, which has a high energy of approximately a few tens to a hundred MeV, provides additional angular momentum to the compound system.
The population ratio of the high-spin isomer ($r^\mathit{is} \equiv b^\mathit{is}/b^\mathrm{tot}$) increases with an increase in the number of neutron evaporation.
As shown in Table~\ref{tab:result}, $r^\mathit{is}$ of $^{106m}$Pd was 15\% for the 0n channel (muon capture of $^{106}$Pd), 33\% for the 2n channel ($^{108}$Pd), and 35\% for the 4n channel ($^{110}$Pd).
A similar trend was found for $^{104m}$Rh and $^{108m}$Rh productions, except for the small $r^\mathit{is}$ value for $^{106}\mathrm{Pd}(\mu^-,2n\nu_\mu){}^{104m}\mathrm{Rh}$ of 1.3\%, the origin of which is not understood.
As the energy of the recoiled neutrino decreases with high multiplicity for neutron emission, the increase in $r^\mathit{is}$ for many neutron evaporations indicates that the recoils of the evaporated neutrons are a major source of angular momentum to the residual nuclei, and the effect from the neutrino recoil is relatively smaller than that of the neutrons.

\begin{table}
    \centering
    \caption{Summary of the production branching ratios for charged particle emission channels.}
    \label{tab:chargedparticle}
    \begin{ruledtabular}
    \begin{tabular}{cccccc}
         Channel & $^{104}$Pd & $^{105}$Pd & $^{106}$Pd & $^{108}$Pd & $^{110}$Pd  \\
         \hline
         1p0n & 0.18(6)\% &           & $<$0.3\%  & $<$1.4\%  & $<$0.23\%\\
         1p1n &           &           &           &           & $<$0.3\% \\
         1p2n &           &           &           & $<$0.3\%  & $<$0.4\% \\
         2p2n & $<$2.5\%  & $<$0.05\% &           & $<$0.24\% & $<$0.26\% \\
    \end{tabular}
    \end{ruledtabular}
\end{table}

The production branching ratio for charged-particle emission channels is generally very small because of the Coulomb barrier, as shown in Table~\ref{tab:chargedparticle}.
The branching ratio of the 1p0n channel for muon capture of $^{104}$Pd (production of $^{103}$Ru) was measured to be $b=0.18(6)$\% in the present experiment.
Several upper limits for one proton emission channels (1p$x$n) were also obtained below 0.3\%, indicating that the proton emission probability was approximately 0.1\%.
Wyttenbach \textit{et al.} investigated the charged-particle emission probabilities for several nuclei and proposed systematics of the branching ratios as a function of the Coulomb barrier height~\cite{Wyttenbach1978-ss}.
The classical Coulomb barrier ($V_C$) is calculated as follows:
\begin{equation}
    V_C = \frac{e^2}{4\pi\epsilon_0}\frac{zZ}{r_0A^{1/3}+\rho_0},
\end{equation}
where $z$ and $Z$ are the charges of the outgoing charged particle and the residual nucleus, respectively; $\frac{e^2}{4\pi\epsilon_0}$ is taken as 1.44~MeV, $r_0$ is 1.35~fm, and $\rho_0$ is 1.2~fm for $\alpha$ particle and 0~fm for proton.
For $^{103}$Ru production from $^{104}$Pd capture, $V_C=9.84$~MeV.
From Wytternbach's systematics, a probability of approximately 0.08\% is predicted, which agrees with our obtained values.
The PHITS calculation predicted proton emission probabilities at around 1--2\%.
Despite the slight overestimation of the excited energy of the compound nucleus by muon capture in the calculation discussed above, the obvious overestimation of the proton emission probability may originate from the underestimation of the surface effects and/or in the evaporation process in GEM, which needs improvement.
No alpha emission channels (2p2n channel) were observed in the present study.
The predicted alpha emission probabilities are approximately 0.01\% from the systematics~\cite{Wyttenbach1978-ss} and 0.03\% from the PHITS calculation, which are below the present detection limits.
Improvements in the experimental method and setup are required to measure the charged particle emission channels, as discussed in the next subsection.

\subsection{In-beam activation method}

We have developed a novel method of in-beam activation to obtain the production branching ratio of muon capture.
As the present study is the first application of this new method, we discussed its features.

The in-beam activation method enables the measurement of most of the activation within a few milliseconds to several hours.
The combined use of classical offline activation with in-beam activation is essential when some of the half-lives of the residual nuclei are extremely long and most of the reaction products can be measured for completeness of the muon-induced nuclear reaction data.

In most situations, the measurement accuracy of the production branching ratio in this experiment was not limited by statistics.
In the activation method, the \textit{absolute} $\beta$-decay branch ($I_\gamma$) for observed $\gamma$ rays must be known.
In the present case, most \textit{relative} uncertainties in the production branching ratio are dominated by those in the \textit{absolute} $\beta$ decay branch ($\Delta I_\gamma^\mathrm{abs}$) of the rhodium isotopes.
The measurement of absolute $I_\gamma$ values at modern radioisotope beam facilities is important for improving the accuracy of data.
Therefore, we presented our experimental observations ($N_\gamma/(\epsilon_\gamma \epsilon_\mathrm{LT})$) separately from the values reported in the literature ($I_\gamma$) in Tables~\ref{tab:108Pd}--\ref{tab:110Pd} for future improvements and reevaluations.
More importantly, the total branching ratios are limited by their absolute uncertainties, the compositions of which are listed in Sect.~\ref{sec:branch}.
Although there is room for improvement, the practical limit of the measurement accuracy using the in-beam activation method might be 5\%.

The sensitivity of the measurement strongly depends on the decay properties of radioactive residual nuclei.
In general, the sensitivity of in-beam activation measurement is higher for radioactive nuclei with short half-lives ($T_{1/2}$), reflecting a high $P_\mathrm{decay}$ value, high $\gamma$-ray intensity ($I_\gamma$), and low $\gamma$-ray energy, reflecting high detection efficiency ($\epsilon_\gamma$).
The production of $^{104m}$Rh and $^{104g}$Rh during the $^{108}$Pd activation is a good example.
A small branching ratio of 0.80(10)\% is obtained for the IT state of $^{104m}$Rh owing to the high detection efficiency for low-energy $\gamma$ ray at 51.5 keV ($\epsilon_\gamma = 3.5$\%) and a large $I_\gamma$ of 48\%.
As the IT state decays to the ground state, the production branching ratio of the ground state should be greater than that of the isomeric state.
We only obtained a detection limit (upper limit) of 13\% for $^{104g}$Rh because of the moderate $\gamma$-ray energy at 555.8 keV ($\epsilon_\gamma = 0.9$\%) and a small $I_\gamma$ of 2.0\%.
The sensitivity of the in-beam activation setup in the present study was approximately 0.1\% for the best cases, for example, $^{107m}$Pd, $^{104m}$Rh, $^{103}$Ru and $^{101}$Rc, and there is room for improvement.
As half of the count rate of the germanium detector is from the environmental background, building up more lead shields around the detector setup will reduce the background.
The use of an anti-Compton shield improves the signal-to-background ratio in the $\gamma$-ray spectrum.
Because some of the $\beta$ rays hit the $\gamma$-ray detector, an anti-$\beta$-ray counter placed in front of the germanium detector will help reduce the background.
Considering the above improvements, the detection limit of the in-beam activation method may be 0.01\% for the best-case scenario.

This method also provides the possibility of extracting half-lives, as demonstrated for the $^{107m}$Rh case in Fig.~\ref{fig:pd107m}.
The half-life of $^{107m}$Rh is known to be $>10$~$\mu$sec and was constrained to be 0.3--10~sec in the present experiment.
The obtained lower limit of 0.3~sec corresponds to the upper limit for extracting the half-life using this method.

The in-beam activation method is applicable only at pulsed muon beam facilities.
The design of the experimental setup is completely different from that of the muonic X-ray and prompt $\gamma$-ray measurements at the pulsed muon facility, for which the high multiplicity of photons at the prompt timing of beam arrival is the main concern in planning the experimental setup.
To avoid pile-up due to the multiple photon detection in a single detector, the detectors are placed sufficiently far from the target or have high granularity, or the muon beam intensity is reduced.
The advantage of the in-beam activation method is using the full capability of muon beam intensity and photon detectors.
The in-beam activation abandoned the measurements of the prompt events in favor of the measurement of the delayed $\gamma$ rays, and the prompt events are eliminated from the analysis as dead times, as explained in Sect.~\ref{sec:analysis}.
Therefore, a large volume detector can be used and placed very close to the target, and the full beam intensity can be accepted as long as the analysis dead time ($T_d$) is shorter enough than the interpulse period.

\section{Conclusion}
\label{sec:conclusion}

In this study, a new methodology called the in-beam activation method, was developed to obtain the radioactivity of short half-lives using the activation method.
The use of the in-beam activation method combined with classical offline activation enables the measurement of most of the radioactivity within a few milliseconds to several years.
As for the first application of the new method, we measured the production branching ratios of muon capture for five palladium isotopes: $^{104,105,106,108,110}$Pd.
The results were compared with model calculations using the PHITS code, which well reproduced the experimental data.
For the first time, this study provides concrete experimental data on the distribution of production branching ratios without any theoretical estimation or assumptions in the interpretation of the data analysis. 

\begin{acknowledgments}
The authors acknowledge the accelerator and technical staff at RAL-ISIS for their support. 
We thank Prof.~Igashira and Prof.~Katabuchi for providing the isotope-enriched palladium targets, Dr.~Y.~Inoue for providing the UT offline setup, and Dr.~S.~Abe for providing the PHITS calculations and discussions.
This work was funded by the ImPACT Program of the Council for Science, Technology, and Innovation (Cabinet Office, Government of Japan), and partially supported by JSPS KAKENHI (Grant Number: JP18J10554).
T.Y.S. acknowledges the support obtained from the ALPS Program at the University of Tokyo.
\end{acknowledgments}

\bibliography{paperpile}

\end{document}